\documentclass[preprint,3p,12pt]{elsarticle}

\usepackage{amssymb}
\usepackage{amsmath}

\usepackage{booktabs}
\usepackage{multicol,multirow}
\usepackage{subfig}
\usepackage{hyperref}
\hypersetup{
colorlinks=true,
linkcolor=red,
filecolor=blue,
citecolor = green,
urlcolor=cyan,
}
\usepackage{tikz}
\usetikzlibrary{positioning, shapes, arrows.meta, fit}
\usepackage{pgfplots}
\pgfplotsset{compat=1.18}

\usepackage[table]{xcolor} 

\journal{}

\begin{document}

\begin{frontmatter}





\title{How Well Do Self-Supervised Speech Models Encode Age and Gender in Children’s Speech? A Layer-Wise Analysis Across Multiple Architectures}


\author[1]{Abhijit Sinha} 
\author[1]{Hemant Kumar Kathania}
\author[1]{Mohit Joshi}
\author[1]{Harishankar Kumar}
\author[2]{Shrikanth Narayanan} 
\author[2]{Sudarsana Reddy Kadiri} 


\affiliation[1]{organization={National Institute of Technology Sikkim}, country={India}}

\affiliation[2]{organization={Signal Analysis and Interpretation Lab (SAIL), University of Southern California}, country={USA}}

\begin{abstract}
Self-supervised learning (SSL) models have become a central component of modern speech processing systems, as they enable the learning of rich acoustic representations without reliance on labeled data. Despite their success on adult speech, it remains unclear how effectively these models capture speaker-related attributes such as age and gender in children’s speech, which differs substantially from adult speech due to ongoing physiological and cognitive development. Higher pitch, increased articulatory variability, and age-dependent acoustic changes make children’s speech a particularly challenging domain. In this work, we present a comprehensive analysis of how age and gender information is encoded across layers of four widely used SSL models: Wav2Vec2, HuBERT, Data2Vec, and WavLM. Layer-wise features are extracted and evaluated using a lightweight convolutional neural network (CNN) on two benchmark children’s speech corpora, PFSTAR and CMU Kids. To analyze feature compactness and redundancy, Principal Component Analysis (PCA) is applied to identify redundancy and highlight the dimensions that contribute most to classification performance. Experimental results show that age- and gender-related information is unevenly distributed across SSL layers, with early to mid-level layers encoding the strongest paralinguistic cues. HuBERT achieves the best overall performance for age classification, while Wav2Vec2 and HuBERT lead gender classification on PFSTAR and CMU Kids, respectively. Beyond single-split evaluation, we further demonstrate that these findings remain stable under speaker-wise cross-validation, layer aggregation, and cross-database evaluation, indicating
robustness to data imbalance and domain mismatch. Finally, we show that reliable age and gender classification is achievable even from short speech segments of 1--3 seconds, highlighting the practical potential of SSL representations for real-world applications involving children’s speech.
\end{abstract}



\begin{keyword}
Children Speech, Self Supervised Learning (SSL), CNN, Dimensionality Reduction.
\end{keyword}

\end{frontmatter}



\section{Introduction}
\label{sec:intro}

Accurately classifying age and gender in children’s speech is essential for advancing child-centered digital technologies that safeguard privacy, ensure safety, and enable personalized user experiences \cite{SAFAVI2018141, NOVOTNY2023104490}. With children spending 4--9 hours daily on digital platforms \href{https://www.aacap.org/AACAP/Families_and_Youth/Facts_for_Families/FFF-Guide/Children-And-Watching-TV-054.aspx}{(AACAP)}, from educational apps to online learning environments, the ability to determine age and gender helps make these interactions both safe and engaging \cite{sarma2020children, mckechnie2018automated}. Reliable classification supports applications such as age-appropriate content curation, adaptive educational materials, and secure access control, creating a protective and personalized environment for young users. Beyond safety, it enhances engagement by tailoring interactions to a child’s developmental stage \cite{10.1121/1.5037614}; for example, age-specific learning recommendations can optimize digital experiences while fostering parental trust. Improved classification accuracy also empowers developers to design services that respect privacy and address children’s unique needs, contributing to a secure, inclusive, and supportive online ecosystem.

However, automatically classifying age and gender from children’s speech presents challenges that differ markedly from adult speech, primarily due to the natural variability in children’s vocal characteristics. Children’s speech often exhibits large pitch fluctuations, shifting formant frequencies, and distinctive pronunciation patterns, all of which evolve with age \cite{koenig2008speech,disfluency}. As children grow, pitch typically lowers, formant frequencies stabilize, and pronunciation begins to resemble that of adults \cite{Vorperian2007VowelAS,lee1999acoustics}. These ongoing developmental changes make it difficult to build models that can accurately classify age and gender across different stages of childhood. In addition, the scarcity of large, well-annotated child-speech datasets further complicates model training, hindering the development of systems that are both accurate and generalizable \cite{yeung2018difficulties,claus2013survey}.

There has been growing interest in predicting age and gender from children’s speech, although this area remains less explored compared to adult speech. Some studies have used raw waveform inputs with deep learning architectures such as time-delay neural networks (TDNNs) and long short-term memory networks (LSTMs) \cite{sarma2020children}. Others have focused on acoustic and prosodic features \cite{li2013automatic}, including pitch and speaking rate, to estimate a child’s age \cite{kumari2024role}. While both traditional feature-based and deep learning approaches have shown promise, they continue to face challenges due to the inherent variability in children’s speech.

Recent work in this area can be broadly grouped into two main approaches: classification and regression. Classification methods are more common and allow researchers to define custom age groups based on the data \cite{sarma2020children, bocklet2010age, muller2007combining, meinedo2011age, safavi2014identification, kaya2017emotion, perez2018children}. For example, some studies treat all children as a single ``childhood'' group \cite{muller2007combining}, while others use separate age groups for each year \cite{sarma2020children}. Popular classification models include Gaussian Mixture Models with Universal Background Models (GMM-UBM), Multi-Layer Perceptrons (MLPs), Support Vector Machines (SVMs), and Deep Neural Networks (DNNs). However, differing age group definitions make it difficult to compare results across studies. In addition, these models often rely on hundreds or thousands of features, which, when combined with small datasets, increases the risk of overfitting and reduces interpretability from a physiological perspective. Regression-based approaches, by contrast, aim to predict continuous age values and are also being explored \cite{spiegl2009analyzing, van2012calibration, kitagishi2020speaker}.

In contrast, age and gender classification in adult speech has been studied far more extensively. Most studies employ standard features such as Mel-frequency cepstral coefficients (MFCCs) \cite{safavi2016speaker}, pitch, and voice quality measures \cite{li2013automatic}. Common classifiers include GMMs, SVMs, and various deep learning models \cite{kwasny2021gender}. More recent work has explored temporal convolutional neural networks (TCNNs) \cite{sanchez2022age} and filter-based architectures such as SincNet with ERB-scale filters \cite{radha2024automatic}. However, because these models are typically trained on adult speech, their performance degrades when applied to children’s speech due to fundamental differences in vocal development.

The emergence of self-supervised learning (SSL) models such as Wav2Vec2 \cite{baevski2020wav2vec}, HuBERT \cite{hsu2021HuBERT}, Data2Vec \cite{baevski2022Data2Vec}, and WavLM \cite{chen2022WavLM} has revolutionized speech processing by leveraging vast amounts of unlabeled data to learn rich, transferable representations. These models have achieved state-of-the-art results across a wide range of speech processing tasks, including speaker recognition, pathology detection, and emotion recognition \cite{Pepino2021EmotionRF,sinha2024effect,grosz2022Wav2Vec2,Gao2023TwostageFO,9747379,Wav2Vec2_speaker,ssl_emotion,ANIDJAR2024124671,novoselov23_interspeech,tirronen2023utilizing,javanmardi2024pre,javanmardi2024exploring}. However, their application to children’s age and gender classification remains underexplored, with most existing work focusing on full-model fine-tuning \cite{kang2023svldl} or multi-layer feature fusion techniques \cite{kitagishi23_interspeech} for adult speech. While effective, these approaches obscure insights into layer-specific contributions, particularly for children’s speech.

Building on our preliminary work \cite{Abhijit_wocci}, which examined different Wav2Vec2 variants, this study significantly broadens the scope by incorporating additional SSL model architectures—HuBERT, Data2Vec, and WavLM alongside Wav2Vec2. We conduct a detailed analysis of how different layers within these models encode speaker-related information, specifically age and gender, in children’s speech. Features from each transformer layer are extracted and used to train a lightweight convolutional neural network (CNN) classifier. By comparing layer-wise performance across models and tasks, we identify which parts of the network are most informative for age and gender prediction. Our findings indicate that lower to mid-level layers, which retain core acoustic cues, often generalize better to children’s speech than deeper layers. This suggests that valuable speaker-related information is embedded early in the model, enabling effective classification for children’s speech without additional fine-tuning.

\begin{figure*}[!b]
    \centering
    \includegraphics[width=15cm]{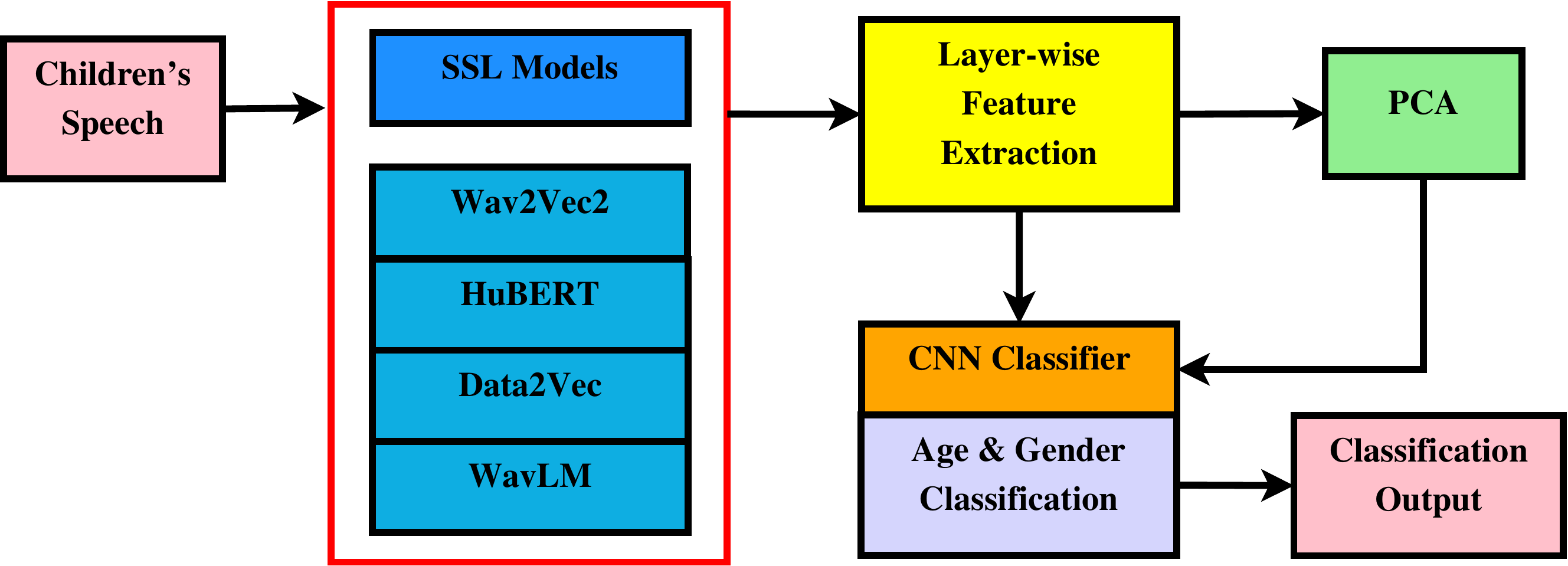} 
    \caption{A block diagram illustrating the proposed framework for classifying age and gender in children using layer-wise features from various SSL models. Dimensionality reduction via PCA is applied to the most effective layer-wise features to enhance performance.}
    \label{fig:block}
\end{figure*}

The main contributions of this study are as follows: 

\begin{itemize}
    \item \textbf{Layer-Wise Evaluation:} We evaluate the layer-wise performance of several SSL models, including Wav2Vec2, HuBERT, Data2Vec, and WavLM, finding that features from early layers significantly outperform those from later layers in age and gender classification accuracy.
    \vspace{5pt}
    \item \textbf{Dimensionality Reduction:} We apply Principal Component Analysis (PCA) to optimize the extracted feature sets, demonstrating that dimensionality reduction improves classification performance. This also reveals that not all SSL features contribute equally to the task.
    \vspace{5pt}
    \item \textbf{Age Group Analysis:} Our analysis shows that classification accuracy improves with age, suggesting that older children’s speech is more distinguishable. This highlights the need for adaptive models that can better handle greater variability in younger children’s speech.
    \vspace{5pt}
    \item {\textbf{Practical Robustness:} We evaluate the robustness of SSL representations through short-segment classification, speaker-wise cross-validation, layer aggregation, and cross-database testing, demonstrating stable performance under limited speech duration and domain mismatch.}
\end{itemize}

The remainder of the paper is organized as follows. Section~\ref{sec:framework} presents the proposed framework for SSL-based feature extraction and analysis. Section~\ref{sec:database} describes the datasets used in this study. Section~\ref{sec:ssl_models} provides an overview of the SSL models employed. Section~\ref{sec:exp_setup} details the experimental setup. Results and discussion are presented in Section~\ref{sec:Result}, and conclusions are drawn in Section~\ref{sec:conclusion}.

\section{Proposed Framework}
\label{sec:framework}

Figure~\ref{fig:block} illustrates the proposed framework for age and gender classification in children’s speech, utilizing layer-wise features extracted from SSL models. The models considered — Wav2Vec2, HuBERT, Data2Vec, and WavLM are pre-trained on large speech corpora and capable of capturing both low-level and high-level features from input speech data. Although originally pretrained and fine-tuned on adult speech corpora, these SSL models were used in a frozen form without any additional fine-tuning for our tasks, allowing us to assess the transferability of their learned representations to age and gender classification on children’s speech.

In the proposed framework, we first extract layer-wise features from the SSL models. These features represent different levels of information, from basic acoustic properties in the early layers to more complex, contextual relationships in the deeper layers. A CNN classifier then processes these layer-wise features, enabling us to evaluate each layer’s contribution to overall classification accuracy. By doing so, we identify which layers provide the most relevant information for the classification task, facilitating better model optimization.

To further enhance classification performance and assess the importance of features for reliable classification, PCA is applied to the most informative layers, for reducing dimensionality and identifying the most relevant features. PCA reduces the dimensionality of the feature sets while preserving the essential characteristics required for accurate classification. {Additionally, to support evaluation under realistic and deployment-oriented conditions, we perform evaluation
on short speech segments to assess performance under limited temporal context, aggregation of information across multiple layers to improve stability, and robustness analysis through cross-validation and cross-database evaluation.}

\section{Database}
\label{sec:database}

This study utilizes two well-known children’s speech datasets: PFSTAR \cite{russell2006pf} and CMU Kids \cite{eskenazi1997cmu}.

\begin{figure}[!h]
    \centering
    \includegraphics[width=\textwidth,height=8cm]{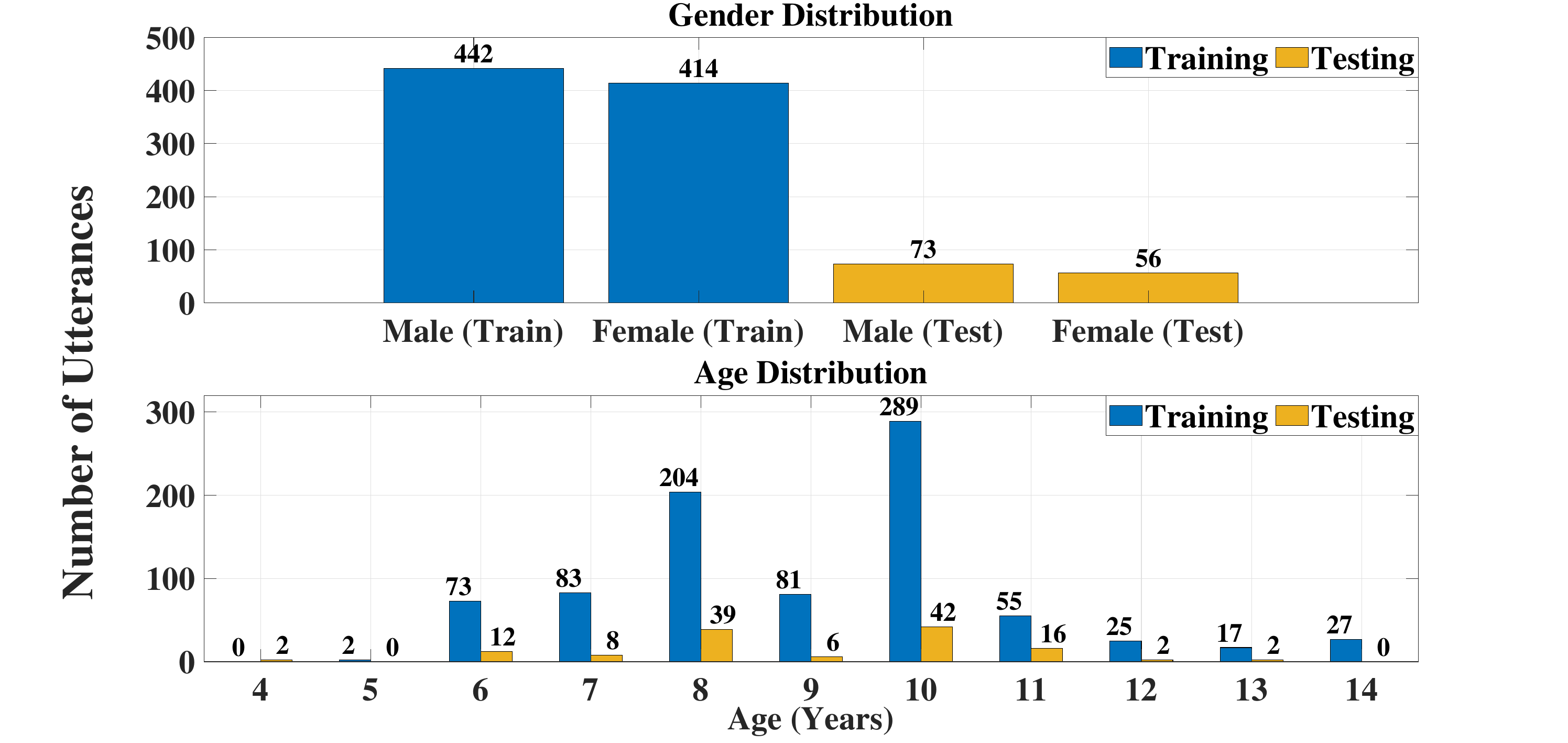} 
    \caption{Age and gender distribution of speakers in the PFSTAR dataset, with training and testing splits presented by gender and age group.}
    \label{fig:data_dist_pfstar}
\end{figure}

The PFSTAR dataset consists of British English speech from children aged 4 to 14 years. It includes 8.3 hours of training speech collected from 122 speakers and 1.1 hours of test speech from 60 speakers. PFSTAR features relatively long utterances, with an average duration of 41.32 seconds, contributing to its higher overall duration despite a smaller number of total utterances. The dataset is organized into 11 age groups and annotated with gender information. {Figure~\ref{fig:data_dist_pfstar} illustrates the age- and gender-wise distribution of speakers in the training and test splits, revealing a naturally imbalanced age distribution, particularly at the younger and older extremes.}

\begin{figure}[!t]
    \centering
    \includegraphics[width=\textwidth,height=8cm]{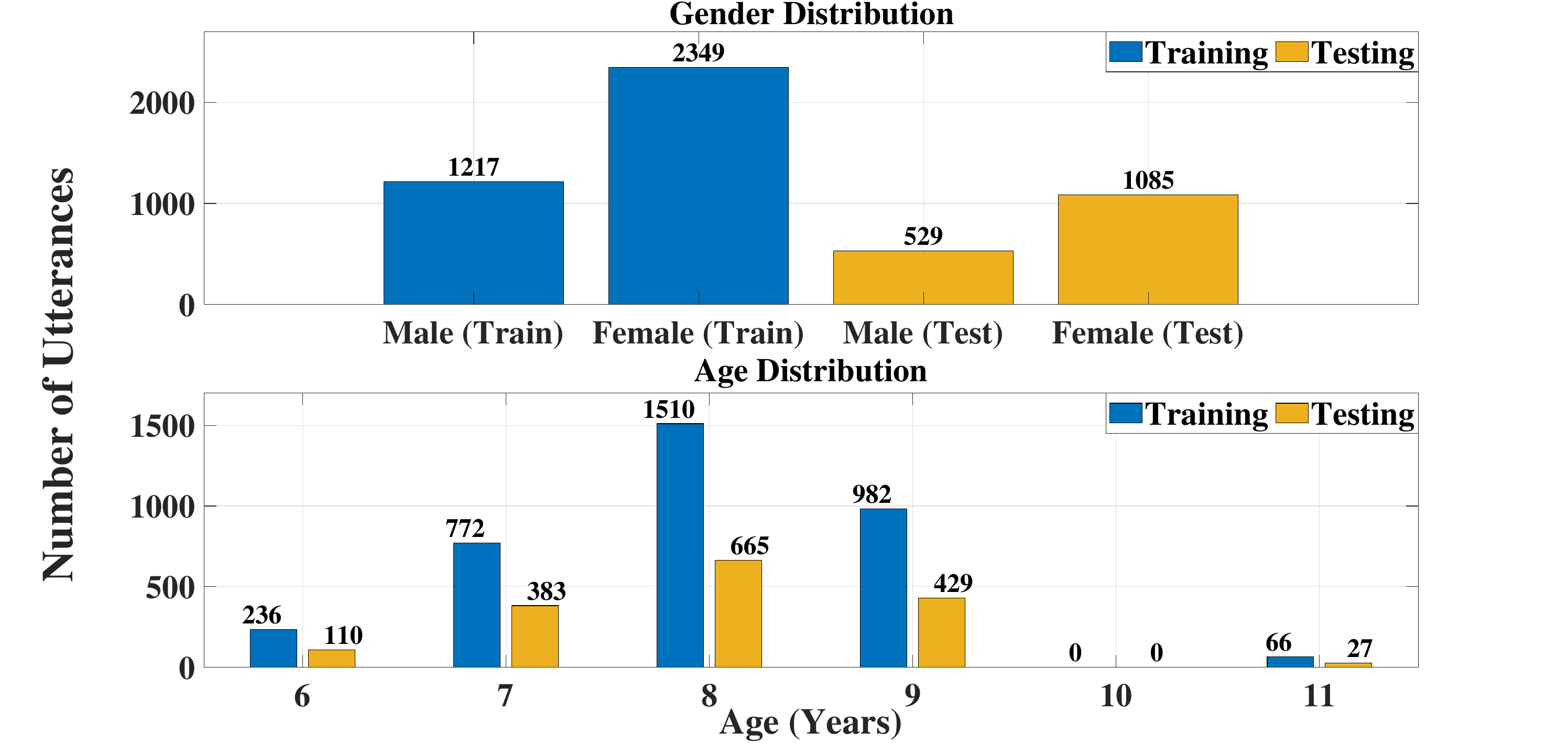}
    \caption{Age and gender distribution of speakers in the CMU Kids dataset, with training and testing splits presented by gender and age group.}
    \label{fig:data_dist_cmu}
\end{figure}

In contrast, the CMU Kids dataset consists of American English speech recorded from
children aged 6 to 11 years. It contains 5,180 utterances from 76 speakers, with 6.3 hours of data used for training and 2.83 hours reserved for testing. Compared to PFSTAR, CMU Kids features much shorter utterances, with an average duration of 6.28 seconds. Age labels are grouped into five classes (6--11 years, excluding age 10), and gender annotations are provided for all speakers. {Figure~\ref{fig:data_dist_cmu} shows the corresponding age- and gender-wise distribution across the training and test sets, indicating a more balanced age structure relative to PFSTAR.}

\begin{table}[!h]
\centering
\caption{Details of the self-supervised learning (SSL) models evaluated in this study. “M” denotes million, “K” denotes thousand, and “h” denotes hours. PT = pretraining; FT = finetuning; Ft. Dim. = feature dimensions for each model.}
\label{tab:ssl_models}
\renewcommand{\arraystretch}{1.2} 
\setlength{\tabcolsep}{6pt} 

\rowcolors{2}{white}{gray!10} 
\begin{tabular}{|l|c|c|c|c|c|}
\hline
\rowcolor{gray!25}
\textbf{Model} & \textbf{Size} & \textbf{PT (h)} & \textbf{FT (h)} & \textbf{Layers} & \textbf{Ft. Dim.} \\
\hline \hline
\multicolumn{6}{|l|}{\textbf{Wav2Vec2}} \\ \hline
base-100h               & 95M   & 960   & 100  & 13 & 768  \\
base-960h               & 95M   & 960   & 960  & 13 & 768  \\
large-960-lv60          & 317M  & 60K   & 960  & 25 & 1024 \\
large-960-lv60-self     & 317M  & 60K   & 960  & 25 & 1024 \\
\hline \hline
HuBERT-large-ls960-ft   & 317M  & 960   & 960  & 25 & 1024 \\
Data2Vec-large          & 317M  & 960   & 960  & 25 & 1024 \\
WavLM-large             & 317M  & 94K   & 960  & 25 & 1024 \\
\hline 
\end{tabular}
\end{table}

\section{Self-Supervised Learning (SSL) Models}
\label{sec:ssl_models}

SSL has emerged as a key approach in speech processing, particularly when labeled data is limited. SSL models, trained on large, unlabeled speech corpora, learn to extract rich, meaningful representations without requiring manual annotations. In this study, we experiment with several state-of-the-art SSL models, including Wav2Vec2 (base-100h, base-960h, large-960h-lv60, large-960h-lv60-self), HuBERT, Data2Vec, and WavLM. These models differ in architecture size, layer depth, and pretraining duration, providing a broad range of features for comparison in the context of age and gender classification from children’s speech. Table~\ref{tab:ssl_models} summarizes the key specifications of the SSL models used in our experiments, including model size, pretraining and fine-tuning durations, number of layers, and feature dimensions.

Each model consists of multiple layers, with each layer capturing different levels of abstraction from the raw speech signal. The layers are indexed from 0 to 24 for the large models and from 0 to 12 for the base models. The zeroth layer corresponds to the output of the convolutional feature encoder, which captures low-level acoustic features such as spectral patterns, phoneme boundaries, and other basic speech characteristics. The remaining layers (1--24 for large models and 1--12 for base models) are transformer layers, designed to capture long-range dependencies and contextual relationships within the speech signal.

The SSL models investigated in this study differ in their architectural designs, influencing how they capture and represent speech features. Wav2Vec2, HuBERT, Data2Vec, and WavLM each employ distinct strategies for feature extraction from raw speech. Wav2Vec2 combines a convolutional feature encoder with transformer layers, enabling it to capture both low-level acoustic features (e.g., spectral patterns and phoneme boundaries) and long-range contextual relationships. In contrast, HuBERT adopts a discrete unit-based pretraining approach, wherein speech segments are clustered into pseudo-labels, and a transformer is trained to predict these units, emphasizing the learning of structured speech patterns. Data2Vec, a more recent model, learns contextualized representations by jointly predicting target states from raw input, allowing it to capture task-relevant information beyond discrete unit prediction. WavLM builds on Wav2Vec2’s architecture by incorporating speaker-aware pretraining, which enhances its robustness to speaker variability––an asset for tasks such as gender classification. These architectural differences result in varied representational characteristics, offering a diverse set of speech features that differ in abstraction level, task specificity, and sensitivity to speaker-related factors.

\section{Experimental Setup}
\label{sec:exp_setup}
In this section, we describe the experimental setup used to evaluate the performance of SSL models for age and gender classification in children’s speech. We outline the baseline method, the process of extracting SSL features, the application of dimensionality reduction via PCA, and the short-segment classification approach. A summary of all conducted experiments are provided in Table~\ref{tab:experiments_summary}.

{
It is important to note that the PFSTAR dataset exhibits a naturally imbalanced age distribution, particularly at the younger (4–6 years) and older (11–14 years) extremes. To ensure that the reported results are reliable under this imbalance and not dependent on a specific data split, we additionally perform speaker-wise cross-validation experiments, as described in Section \ref{sec:robustness}.

}

\subsection{Baseline}

For baseline comparison, we utilize traditional MFCCs, specifically focusing on 26-dimensional features. MFCCs are a well-established choice in speech processing due to their ability to capture critical acoustic properties such as spectral shape and formant structures. While we also evaluated 13- and 39-dimensional MFCC configurations, the 26-dimensional features consistently yielded the highest classification accuracy. These features are extracted directly from speech signals without additional preprocessing and serve as a benchmark to assess the effectiveness of SSL-based representations for age and gender classification.

For all experiments, we employed the framework illustrated in Figure~\ref{fig:block}. This approach involves extracting MFCC features and layer-wise SSL features and subsequently processing these features through a 1D CNN for classification. The CNN consists of three convolutional layers with increasing numbers of filters: 32, 64, and 128, each using a kernel size of 5. ReLU activation functions are applied after each layer to introduce non-linearity, enabling the network to learn complex patterns. Additionally, batch normalization is used after each convolutional layer to stabilize the learning process and improve overall training efficiency and performance.

{
The choice of a shallow 1D CNN classifier is intentional and aligned with the analytical goals of this study. Our objective is not to optimize downstream classification performance, but to examine how age- and gender-related information is distributed across different layers of SSL models. A lightweight CNN provides sufficient capacity to model local temporal patterns in SSL embeddings while avoiding the use of highly expressive architectures. This ensures that the observed performance trends primarily reflect the representational properties of the SSL features themselves rather than the modeling power of the classifier.

}

\subsection{SSL Features}

We experiment with several state-of-the-art SSL models, including Wav2Vec2 (base and large variants), HuBERT, Data2Vec, and WavLM for extracting frame-level representations from children’s speech. Each model consists of a convolutional feature encoder followed by a stack of transformer layers (e.g., 12 layers for base models, 24 layers for large models). We extract features from both the convolutional output (before the transformer) and from each transformer layer to evaluate their effectiveness for age and gender classification. The convolutional layer captures low-level acoustic patterns, while the transformer layers progressively encode higher-level information. The extracted features are used as input to a CNN-based classifier, and performance is compared against a baseline using traditional MFCC features. Our goal is to assess whether SSL-based representations offer more robust and discriminative features for children’s speech and to identify which specific layers contribute most to classification accuracy.

\subsection{Dimensionality Reduction}

To assess the relevance of the extracted SSL features, we apply Principal Component Analysis (PCA) for dimensionality reduction. PCA identifies the principal components that account for the maximum variance in the feature set and projects the original SSL features into a lower-dimensional space. This process helps determine whether all features contribute meaningfully to the classification task or if some are redundant, improving both computational efficiency and model generalization. We experimented with reducing the feature dimensions from 512 down to 64 in steps of 64, and further down to 32, to identify the optimal feature space for age and gender classification.

{
We emphasize that PCA is used here as a deterministic, layer-independent tool to assess dimensional sufficiency within fixed SSL representations, rather than as a trainable projection for learning task-specific features.

}

\subsection{Age Group-Wise Analysis}

In addition to evaluating the overall classification performance, we performed an age and gender group-wise analysis to examine how well the model detects age and gender across different subgroups. For the PFSTAR dataset, we divided the data into three age groups: 4--6, 7--9, and 10--13 years. For the CMU Kids dataset, we used two age groups: 6--8 and 9--11 years. Within each age group, we assessed the model’s ability to classify both age and gender. This analysis helps us understand if the model’s performance differs across various age ranges and gender categories, providing insights into the SSL features’ sensitivity to age-dependent and gender-specific speech characteristics.

\subsection{Short-Segment Classification}

In addition to using the original full-length test samples, we performed segmentation to better simulate real-world scenarios where only brief speech inputs may be available. Voice Activity Detection (VAD) was first applied to remove silence from the speech recordings. The resulting speech segments were then divided into fixed durations of 1 second, 2 seconds, and 3 seconds. This segmentation approach allows us to analyze how the length of speech input affects classification performance. Such conditions are common in practical applications like interactive voice assistants, child monitoring systems, or educational tools, where users may speak in short phrases or utterances. By evaluating different segment lengths, we aim to understand the model’s ability to adapt to varying input durations and its effectiveness in real-life use cases.

{
VAD ensures that all segments contain voiced speech, and while shorter segments (1-second) may contain partial words, the 2- and 3-second segments typically include complete words or multiple syllables. The analysis is intentionally focused on acoustic rather than lexical cues, which are known to dominate age and gender perception.

}

\begin{table*}[!htbp]
\centering
\caption{Summary of experimental setups and corresponding results tables. Each row outlines the experiment type, its objective, and the section where it is described, with direct references to the tables presenting the results for both PFSTAR and CMU Kids datasets.}
\label{tab:experiments_summary}
\renewcommand{\arraystretch}{1.2} 
\setlength{\tabcolsep}{5pt} 
\resizebox{\textwidth}{!}{
\begin{tabular}{|c|p{6cm}|p{10.5cm}|p{4cm}|}
\hline
\rowcolor{gray!25}
\textbf{Section} & \textbf{Experiment} & \textbf{Description} & \textbf{Results} \\
\hline
\ref{subsec:baseline} & Baseline (MFCC) & Baseline performance on the PFSTAR dataset and CMU Kids dataset using traditional MFCC features. & Tables \ref{tab:baseline_layerwise_pfstar}, \ref{tab:baseline_layerwise_cmu} \\
\hline
\ref{subsec: Layer wise performance} & Layer-wise analysis & Layer-specific classification performance on PFSTAR and CMU Kids datasets to identify the most effective layers from various SSL models. & Tables \ref{tab:baseline_layerwise_pfstar}, \ref{tab:baseline_layerwise_cmu} \\
\hline
\ref{subsec: Dimensionality Reduction using PCA} & Dimensionality reduction (PCA) & PCA applied to PFSTAR and CMU Kids dataset features to reduce dimensionality and evaluate its effect on classification accuracy. & Tables \ref{tab:PCA_PFSTAR}, \ref{tab:PCA_CMU} \\
\hline
\ref{subsec:age_group_analysis} & Age group analysis & Classification accuracy across age groups for PFSTAR and CMU Kids datasets to study performance variations due to developmental changes. & Tables \ref{tab:age_group_pfstar}, \ref{tab:age_group_cmu} \\
\hline
\ref{subsec: Audio Segmentation} & Short-segment classification & PFSTAR and CMU Kids test sets segmented into 1s, 2s, and 3s chunks to analyze the impact of segment duration on performance. & Tables \ref{tab:chunk_pfstar}, \ref{tab:chunk_cmu} \\
\hline
\textcolor{red}{\ref{sec:robustness}} &
\textcolor{red}{Robustness and Generalization of SSL Representations} &
\textcolor{red}{Evaluated cross-database transfer, layer aggregation strategies, and k-fold cross-validations.} &
\textcolor{red}{Tables \ref{tab:cross-db-gender}, \ref{tab:crossdb-gender-HuBERT3}, \ref{tab:HuBERT6-gender-crossdb},
\ref{tab:pfstar-agg}, \ref{tab:cmu-agg}, \ref{tab:pfstar-baseline}, \ref{tab:cmu-baseline},
\ref{tab:pfstar-age-gender}, \ref{tab:cmukids-age-gender}} \\
\hline
\end{tabular}}
\end{table*}

\section{Results and Discussion}
\label{sec:Result}
\subsection{Baseline}
\label{subsec:baseline}

To provide a baseline for comparison, we evaluated the performance of a traditional MFCC-based model on the PFSTAR and CMU Kids datasets for both age and gender classification tasks. While MFCCs are a standard choice in speech processing, our results highlight their limitations in capturing the complex, age-dependent variations present in children’s speech. We report key performance metrics accuracy, precision, recall, and F1 score which serve as reference points for assessing improvements achieved through SSL-based feature representations.

On the PFSTAR dataset, baseline age classification achieved an accuracy of 80.92\% (precision: 0.82, recall: 0.81, F1: 0.80), while gender classification yielded higher performance at 87.63\% accuracy (precision: 0.90, recall: 0.88, F1: 0.88). These results highlight MFCCs’ limitations in encoding age-specific vocal traits, which exhibit greater intra-class variability due to developmental changes in vocal tract length and articulation stability. The CMU Kids dataset demonstrated improved age classification accuracy of 89.97\% (precision: 0.90, recall: 0.89, F1: 0.89), attributable to its narrower age range (6--11 years, 5 classes excluding age 10), reducing inter-speaker variability compared to PFSTAR’s broader 4--14 years (11 classes) span. Gender classification on CMU Kids achieved 88.41\% accuracy (precision: 0.89, recall: 0.88, F1: 0.88), closely matching PFSTAR’s results. This consistency across datasets suggests that gender-discriminative features such as pitch and formant spacing are more robust to recording condition variations than age-related acoustic markers.  

These baseline results, summarized in Tables~\ref{tab:baseline_layerwise_pfstar} and \ref{tab:baseline_layerwise_cmu}, underscore the limitations of MFCC features for children’s speech analysis. While MFCCs capture fundamental spectral properties, their inability to model longitudinal developmental patterns and articulatory cues results in suboptimal age classification performance. This gap motivates our exploration of SSL-based feature extraction, which leverages self-supervised pretraining to encode richer, context-aware representations of pediatric speech.

\subsection{Layer wise performance}
\label{subsec: Layer wise performance}

\begin{figure*}[!h]
    \centering
    \includegraphics[width=\textwidth]{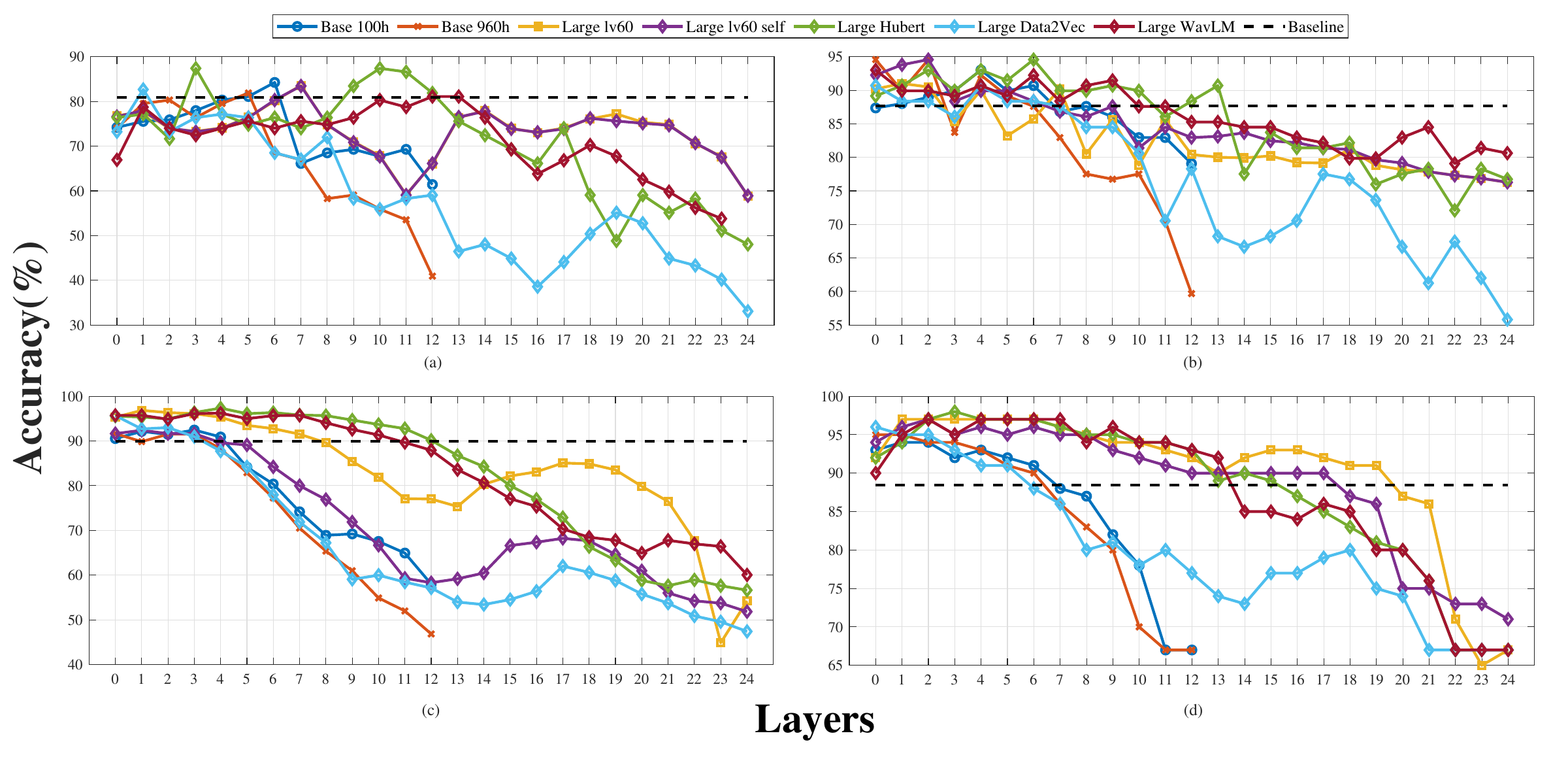} 
    \caption{Layer-wise classification performance for age and gender prediction across four Wav2Vec2 models. Subplots show: (a) age classification accuracies on the PFSTAR dataset, (b) gender classification accuracies on the PFSTAR dataset, (c) age classification accuracies on the CMU Kids dataset, and (d) gender classification accuracies on the CMU Kids dataset.}
    \label{fig:layer_wise}
     \vspace{-3pt}
\end{figure*}

\begin{table}[!h]
\centering
\caption{Performance metrics — accuracy (A), precision (P), recall (R), and F1 score (F1) — for the best-performing layers of the Wav2Vec2, HuBERT, Data2Vec, and WavLM models in age and gender classification tasks on the PFSTAR dataset.}
\label{tab:baseline_layerwise_pfstar}
    \resizebox{16cm}{!}{
\begin{tabular}{|c|c|c|cccc|c|cccc|}
\hline
\textbf{Model} & \textbf{SSL Variant} & \textbf{Best Layer} & \multicolumn{4}{c|}{\textbf{AGE}} & \textbf{Best Layer} & \multicolumn{4}{c|}{\textbf{GENDER}} \\
\cline{4-7} \cline{9-12}
 & & & \textbf{A (\%)} & \textbf{P} & \textbf{R} & \textbf{F1} & & \textbf{A (\%)} & \textbf{P} & \textbf{R} & \textbf{F1} \\
\hline
\textbf{Baseline} & MFCC & - & 80.92 & 0.82 & 0.81 & 0.80 & - & 87.63 & 0.90 & 0.88 & 0.88 \\
\hline
\multirow{4}{*}{\textbf{Wav2Vec2}} 
& base-100h            & 6 & 84.25 & 0.86 & 0.84 & 0.83 & 4 & 93.02 & 0.93 & 0.93 & 0.92 \\
& base-960h            & 5 & 81.89 & 0.84 & 0.82 & 0.81 & 2 & 94.57 & 0.95 & 0.94 & 0.93 \\
& large-960h-lv60      & 7 & 83.59 & 0.84 & 0.83 & 0.83 & 1 & 91.45 & 0.93 & 0.92 & 0.90 \\
& large-960h-lv60-self & 7 & 83.46 & 0.85 & 0.84 & 0.83 & 2 & \textbf{94.57} & 0.95 & 0.94 & 0.94 \\
\hline
\textbf{HuBERT} & HuBERT-large-ls960-ft & 3 & \textbf{87.40} & 0.88 & 0.87 & 0.87 & 6 & \textbf{94.57} & 0.96 & 0.94 & 0.93 \\
\hline
\textbf{Data2Vec} & Data2Vec-large & 1 & 82.68 & 0.86 & 0.83 & 0.83 & 0 & 90.70 & 0.91 & 0.91 & 0.91 \\
\hline
\textbf{WavLM} & WavLM-large & 13 & 81.10 & 0.84 & 0.81 & 0.81 & 0 & 93.02 & 0.94 & 0.94 & 0.93 \\
\hline
\end{tabular}}
\end{table}

\begin{table}[!h]
\centering
\caption{Performance metrics — accuracy (A), precision (P), recall (R), and F1 score (F1) — for the best-performing layers of the Wav2Vec2, HuBERT, Data2Vec, and WavLM models in age and gender classification tasks on the CMU Kids dataset.}
    \resizebox{16cm}{!}{
\begin{tabular}{|c|c|c|cccc|c|cccc|}
\hline
\multirow{2}{*}{\textbf{Model}} & \multirow{2}{*}{\textbf{SSL Model}} & \multirow{2}{*}{\textbf{Best Layer}} & \multicolumn{4}{c|}{\textbf{AGE}} & \multirow{2}{*}{\textbf{Best Layer}} & \multicolumn{4}{c|}{\textbf{GENDER}} \\
\cline{4-7} \cline{9-12}
& & & \multicolumn{1}{c|}{\textbf{A (\%)}} & \multicolumn{1}{c|}{\textbf{P}} & \multicolumn{1}{c|}{\textbf{R}} & \textbf{F1} & & \multicolumn{1}{c|}{\textbf{A (\%)}} & \multicolumn{1}{c|}{\textbf{P}} & \multicolumn{1}{c|}{\textbf{R}} & \textbf{F1} \\
\hline
\textbf{Baseline} & MFCC & - & \multicolumn{1}{c|}{89.97} & \multicolumn{1}{c|}{0.90} & \multicolumn{1}{c|}{0.89} & 0.89 & - & \multicolumn{1}{c|}{88.41} & \multicolumn{1}{c|}{0.89} & \multicolumn{1}{c|}{0.88} & 0.88 \\
\hline
\multirow{4}{*}{\textbf{Wav2Vec2}} 
& base-100h & 1 & \multicolumn{1}{c|}{92.13} & \multicolumn{1}{c|}{0.92} & \multicolumn{1}{c|}{0.92} & 0.92 & 2 & \multicolumn{1}{c|}{93.78} & \multicolumn{1}{c|}{0.94} & \multicolumn{1}{c|}{0.94} & 0.94 \\
& base-960h & 0 & \multicolumn{1}{c|}{91.63} & \multicolumn{1}{c|}{0.90} & \multicolumn{1}{c|}{0.90} & 0.90 & 1 & \multicolumn{1}{c|}{94.96} & \multicolumn{1}{c|}{0.95} & \multicolumn{1}{c|}{0.95} & 0.95 \\
& large-960h-lv60 & 1 & \multicolumn{1}{c|}{96.84} & \multicolumn{1}{c|}{0.97} & \multicolumn{1}{c|}{0.97} & 0.97 & 2 & \multicolumn{1}{c|}{96.68} & \multicolumn{1}{c|}{0.97} & \multicolumn{1}{c|}{0.97} & 0.97 \\
& large-960h-lv60-self & 1 & \multicolumn{1}{c|}{92.37} & \multicolumn{1}{c|}{0.92} & \multicolumn{1}{c|}{0.92} & 0.92 & 2 & \multicolumn{1}{c|}{96.53} & \multicolumn{1}{c|}{0.97} & \multicolumn{1}{c|}{0.97} & 0.97 \\
\hline
\textbf{HuBERT} & HuBERT-large-ls960-ft & 4 & \multicolumn{1}{c|}{\textbf{97.33}} & \multicolumn{1}{c|}{0.97} & \multicolumn{1}{c|}{0.97} & 0.97 & 3 & \multicolumn{1}{c|}{\textbf{98.00}} & \multicolumn{1}{c|}{0.98} & \multicolumn{1}{c|}{0.98} & 0.98 \\
\hline
\textbf{Data2Vec} & Data2Vec-large & 0 & \multicolumn{1}{c|}{95.66} & \multicolumn{1}{c|}{0.96} & \multicolumn{1}{c|}{0.96} & 0.96 & 0 & \multicolumn{1}{c|}{96.00} & \multicolumn{1}{c|}{0.96} & \multicolumn{1}{c|}{0.96} & 0.95 \\
\hline
\textbf{WavLM} & WavLM-large & 4 & \multicolumn{1}{c|}{96.22} & \multicolumn{1}{c|}{0.96} & \multicolumn{1}{c|}{0.96} & 0.96 & 4 & \multicolumn{1}{c|}{97.12} & \multicolumn{1}{c|}{0.97} & \multicolumn{1}{c|}{0.97} & 0.97 \\
\hline
\end{tabular}}
\label{tab:baseline_layerwise_cmu}
\end{table}
To further evaluate the effectiveness of SSL features for age and gender classification, we conducted a layer-wise analysis using Wav2Vec2, HuBERT, Data2Vec, and WavLM models on the PFSTAR and CMU Kids datasets. This analysis aims to determine how features extracted from different layers contribute to classification performance, with a particular focus on the initial layers, which are expected to capture acoustically discriminative features crucial for these tasks.

Figure~\ref{fig:layer_wise} provides a graphical comparison of the baseline and layer-wise performance for both age and gender classification. These graphs offer visual evidence of the substantial improvements achieved by SSL models, particularly from the initial layers. The figures highlight the significant contribution of these layers to classification performance, underlining the importance of utilizing SSL features for such tasks. Performance gains on CMU Kids reflect the impact of its controlled recording conditions and narrower age range (6--11 years). However, gender classification showed minimal dataset variance. Figure~\ref{fig:layer_wise} illustrates these trends, with early-layer peaks for both tasks. The results, summarized in Tables~\ref{tab:baseline_layerwise_pfstar} and \ref{tab:baseline_layerwise_cmu}, present performance metrics — accuracy, precision, recall, and F1 score of the best-performing layers for each model, in comparison with baseline results using 26-dimensional MFCC features.

The results consistently demonstrate that the initial layers of the SSL models outperform deeper layers for both age and gender classification. This trend can be attributed to the nature of the features captured by these layers. The initial layers extract low-level acoustic features, which are particularly significant for distinguishing subtle variations in vocal characteristics. These low-level features are essential for tasks like age and gender classification, where vocal timbre and pitch variations are key discriminators. In contrast, deeper layers tend to focus more on high-level semantic content, which is less relevant for classification tasks dependent on these acoustic traits. As a result, deeper layers contribute less effectively to age and gender classification.

For age classification on the PFSTAR dataset, the best-performing layer from the HuBERT large model achieved an accuracy of 87.40\% (precision: 0.88, recall: 0.87, F1: 0.87), representing a significant improvement over the MFCC baseline. Similarly, for gender classification, the Wav2Vec2 large-960h-lv60-self model achieved the best performance with an accuracy of 94.57\% (precision: 0.95, recall: 0.94, F1: 0.94), marking a notable increase from the baseline. These improvements are further supported by enhanced precision, recall, and F1 scores for the best layers, which consistently outperform the baseline across all models. For the CMU Kids dataset, the HuBERT large model outperformed other SSL models for both tasks. For age classification, layer 4 demonstrated an accuracy of 97.33\% (precision: 0.97, recall: 0.97, F1: 0.97), a significant improvement over the MFCC baseline. Similarly, layer 3 achieved an accuracy of 98.00\% (precision: 0.98, recall: 0.98, F1: 0.98) for gender classification.

This layer-wise analysis confirms that for age and gender classification in children’s speech, selecting features from the optimal layers of pre-trained SSL models is crucial. While deeper layers provide richer representations that may be beneficial for other speech processing tasks, the initial layers’ focus on acoustically discriminative features proves to be more effective for tasks like age and gender classification. These findings not only enhance our understanding of SSL model feature contributions but also offer valuable insights for future research aimed at optimizing SSL models for similar speech classification tasks.

Our layer-wise significance analysis, comparing all SSL models to the 26-dimensional MFCC feature baseline, reveals that larger SSL architectures exhibit statistically significant differences \textbf{($p \ll 0.05$)} in both age and gender classification, outperforming smaller/base models in capturing task-specific hierarchical representations.

\subsection{Dimensionality Reduction}
\label{subsec: Dimensionality Reduction using PCA}

\begin{figure*}[!t]
    \centering
    \includegraphics[width=\textwidth]{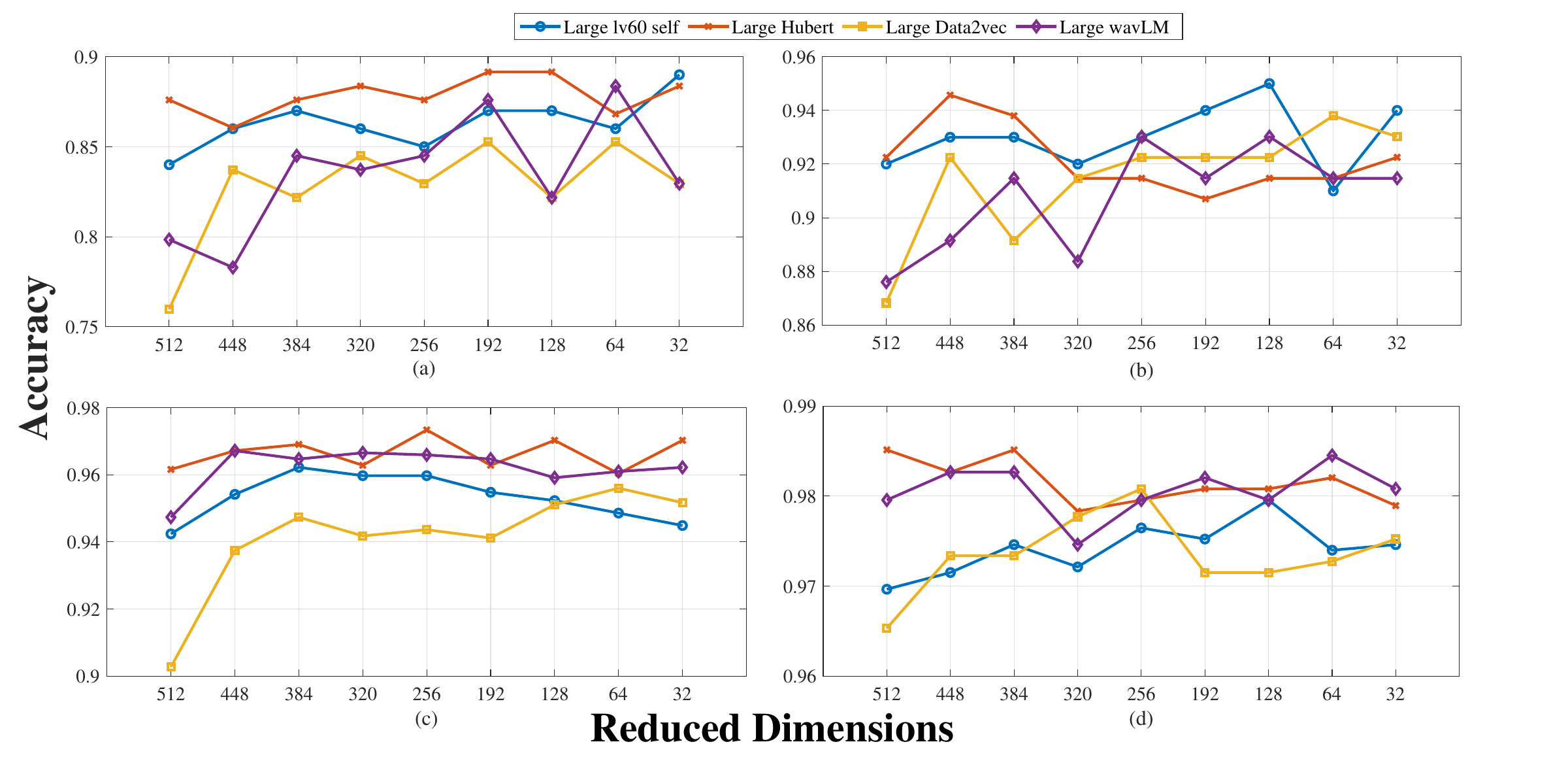} 
    \caption{Effect of PCA-based dimensionality reduction on classification accuracy for age and gender prediction using features extracted from the top-performing layers of large SSL speech models (Wav2Vec2, HuBERT, Data2Vec, and WavLM). The plots illustrate: (a) age classification on the PFSTAR dataset, (b) gender classification on the PFSTAR dataset, (c) age classification on the CMU Kids dataset, and (d) gender classification on the CMU Kids dataset. Each subplot compares accuracy across different numbers of retained PCA components.}
    \label{fig:pca}
     \vspace{-15pt}
\end{figure*}
To evaluate the impact of feature dimensionality on classification performance, we applied PCA to the features extracted from the top-performing layers of the Wav2Vec2, HuBERT, Data2Vec, and WavLM models for age and gender classification. The primary objective of this analysis was to assess whether the full set of features is required for effective classification or if a reduced feature set could achieve comparable performance. The feature dimensions were systematically reduced in steps, ranging from the original number of features down to a minimum of 64 dimensions.

Figure~\ref{fig:pca} provides a graphical comparison of the PCA results for both age and gender classification across all models and datasets. These figures clearly illustrate the performance of each model with reduced feature dimensions, showing that even with significant reduction, classification performance remains robust, confirming that not all features are necessary for optimal classification. The results, summarized in Tables~\ref{tab:PCA_PFSTAR} and \ref{tab:PCA_CMU}, provide insight into how the number of features affects classification accuracy as well as precision, recall, and F1 score.

\begin{table}[t]
    \centering
    \footnotesize
    \caption{Performance metrics — accuracy (A), precision (P), recall (R), and F1 score (F1) — for the best-performing layers of various SSL models in age and gender classification on the PFSTAR dataset, with dimensionality reduction applied using PCA.}
    \resizebox{12cm}{!}{
    \begin{tabular}{|c|c|c|c|c|c|}
        \hline
        \multicolumn{6}{|c|}{\textbf{Age}} \\
        \hline
        \textbf{SSL} & \textbf{Reduced Feature} & {\textbf{A (\%)}} & {\textbf{P}} & {\textbf{R}} & {\textbf{F1}} \\ 
        \textbf{Model} & \textbf{Dimension} & & & & \\
        \hline
        base-100h & 320 & 86.05& 0.87& 0.86& 0.86        \\
        \hline
        base-960h & 384 & 83.72& 0.85& 0.84& 0.82        \\
        \hline
        large-960h-lv60 & 256& 84.8& 0.85& 0.84& 0.83\\
        \hline
        large-960h-lv60-self & 384 & 85.27& 0.87& 0.85& 0.84        \\ \hline 
 HuBERT-large-ls960-ft& 320& \textbf{89.15}& 0.89& 0.89&0.88\\ \hline 
 Data2Vec-large& 320& 85.27& 0.86& 0.85&0.85\\ \hline 
 WavLM-large & 448& 88.37& 0.89& 0.88&0.88\\
        \hline
        \multicolumn{6}{|c|}{\textbf{Gender}} \\
        \hline
        base-100h & 64 & 93.80& 0.94& 0.93& 0.94        \\
        \hline
        base-960h & 384 & 93.80& 0.95& 0.94& 0.94        \\
        \hline
        large-960h-lv60 &320& 92.75& 0.94& 0.92& 0.91\\
        \hline
        large-960h-lv60-self &384 & \textbf{95.00}& 0.95& 0.95& 0.95        \\
        \hline
 HuBERT-large-ls960-ft& 64& 94.57& 0.95& 0.95&0.95\\ \hline 
 Data2Vec-large& 448& 93.80& 0.94& 0.94&0.93\\ \hline 
 WavLM-large & 256& 93.02& 0.93& 0.93&0.93\\ \hline
    \end{tabular}}
    \label{tab:PCA_PFSTAR}
\end{table}

For the PFSTAR dataset (Table~\ref{tab:PCA_PFSTAR}), we observe that significant performance improvements can be achieved with dimensionality reduction. For instance, the best-performing model, HuBERT large, achieved an accuracy of 89.15\% for age classification when reduced to 320 features, marking a substantial increase from the baseline of 80.92\%. Similarly, for gender classification, the Wav2Vec2-base-100h model with only 64 features maintained a high accuracy of 93.80\%, demonstrating that the optimal feature set does not require the full dimensionality of the original features. The Wav2Vec2-large-960h-lv60-self model achieved the highest accuracy of 95.00\% (with 384 features) for gender classification. These results suggest that substantial feature reduction can maintain, and in some cases improve, classification performance, enabling more efficient model deployment without significant loss of accuracy.

In contrast, the results from the CMU Kids dataset (Table~\ref{tab:PCA_CMU}) further reinforce the notion that not all features are necessary for effective classification. For age classification, the Wav2Vec2-large-960h-lv60 model demonstrated an accuracy of 97.14\% with 256 features, while the HuBERT model, with the same number of features, achieved an accuracy of 97.34\%. For gender classification, HuBERT large (with only 32 features) achieved the highest accuracy of 98.51\%, surpassing baseline results with a more compact feature representation.

\begin{table}[t]
    \centering
    \footnotesize
    \caption{Performance metrics — accuracy (A), precision (P), recall (R), and F1 score (F1) — for the best-performing layers of various SSL models in age and gender classification on the CMU Kids dataset, with dimensionality reduction applied using PCA.}
    \resizebox{12cm}{!}{
    \begin{tabular}{|c|c|c|c|c|c|}
        \hline
        \multicolumn{6}{|c|}{\textbf{Age}} \\
        \hline
        \textbf{SSL} & \textbf{Reduced Feature}& \textbf{A (\%)} & \textbf{P} & \textbf{R} & \textbf{F1} \\ 
        \textbf{Model} & \textbf{Dimension}& & & & \\
        \hline
        base-100h & 192                                & 93.18& 0.93& 0.93& 0.93        \\
        \hline
        base-960h & 192                                & 93.43& 0.93& 0.93& 0.93        \\
        \hline
        large-960h-lv60 & 256                                & 97.14& 0.97& 0.97& 0.97        \\
        \hline
        large-960h-lv60-self & 128                                & 96.84& 0.97& 0.97& 0.97        \\ \hline 
 HuBERT-large-ls960-ft& 256& \textbf{97.34}& 0.97& 0.97&0.97\\ \hline 
 Data2Vec-large& 448& 95.60& 0.96& 0.96&0.95\\ \hline 
 wavllm-large & 64& 96.71& 0.97& 0.97&0.97\\
        \hline
        \multicolumn{6}{|c|}{\textbf{Gender}} \\
        \hline
        base-100h & 64                                 & 96.22& 0.96& 0.96& 0.96        \\
        \hline
        base-960h & 256                                & 96.71& 0.97& 0.97& 0.97        \\
        \hline
        large-960h-lv60 & 64                                 & 98.20& 0.98& 0.98& 0.98        \\
        \hline
        large-960h-lv60-self & 384                                & 97.95& 0.98& 0.98& 0.98        \\
        \hline
 HuBERT-large-ls960-ft& 32& \textbf{98.51}& 0.99& 0.98&0.98\\ \hline 
 Data2Vec-large& 256& 98.07& 0.98& 0.98&0.98\\ \hline 
 WavLM-large & 448& 98.45& 0.98& 0.98&0.98\\ \hline
    \end{tabular}}
    \label{tab:PCA_CMU}
\end{table}

\begin{figure*}[!b]
    \centering
    \subfloat[]
    {\includegraphics[width=0.23\textwidth]{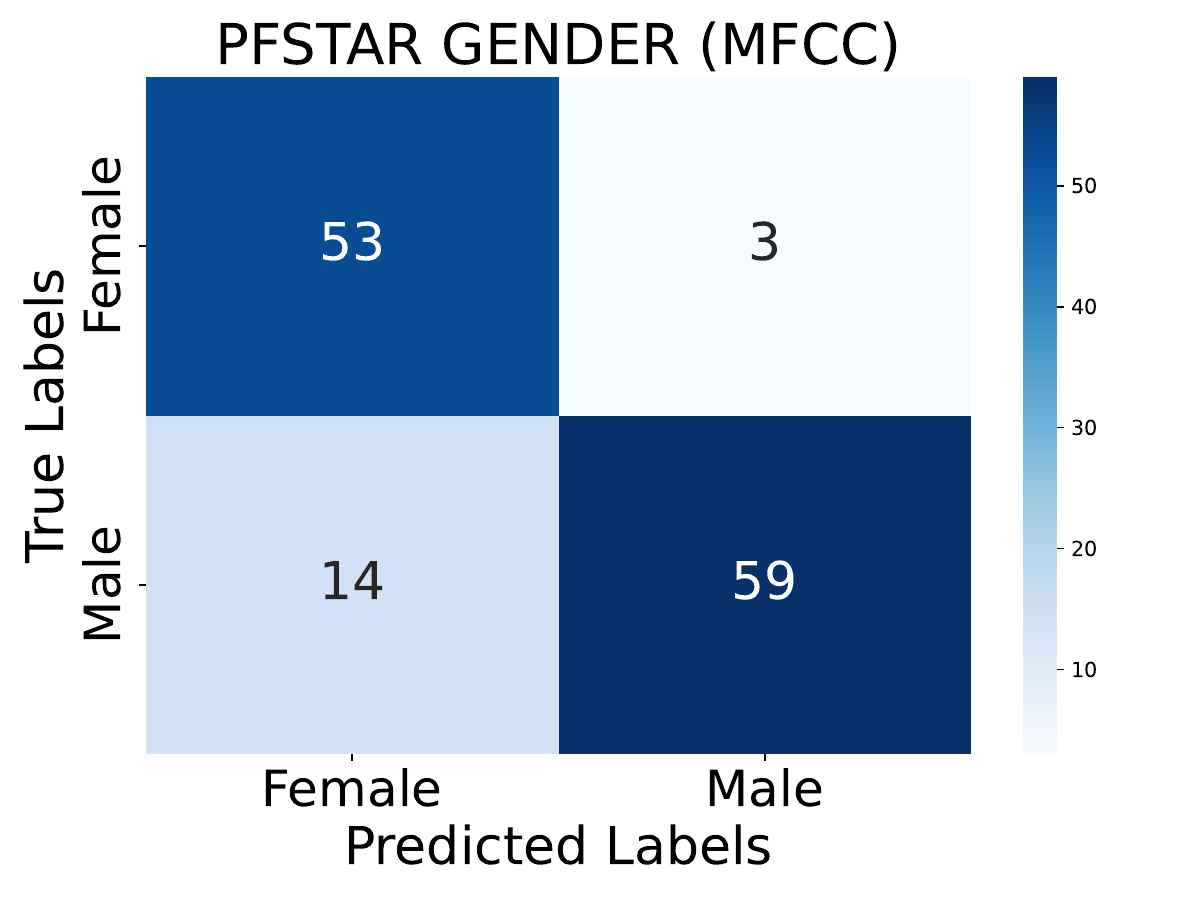}}
    \subfloat[]
    {\includegraphics[width=0.23\textwidth]{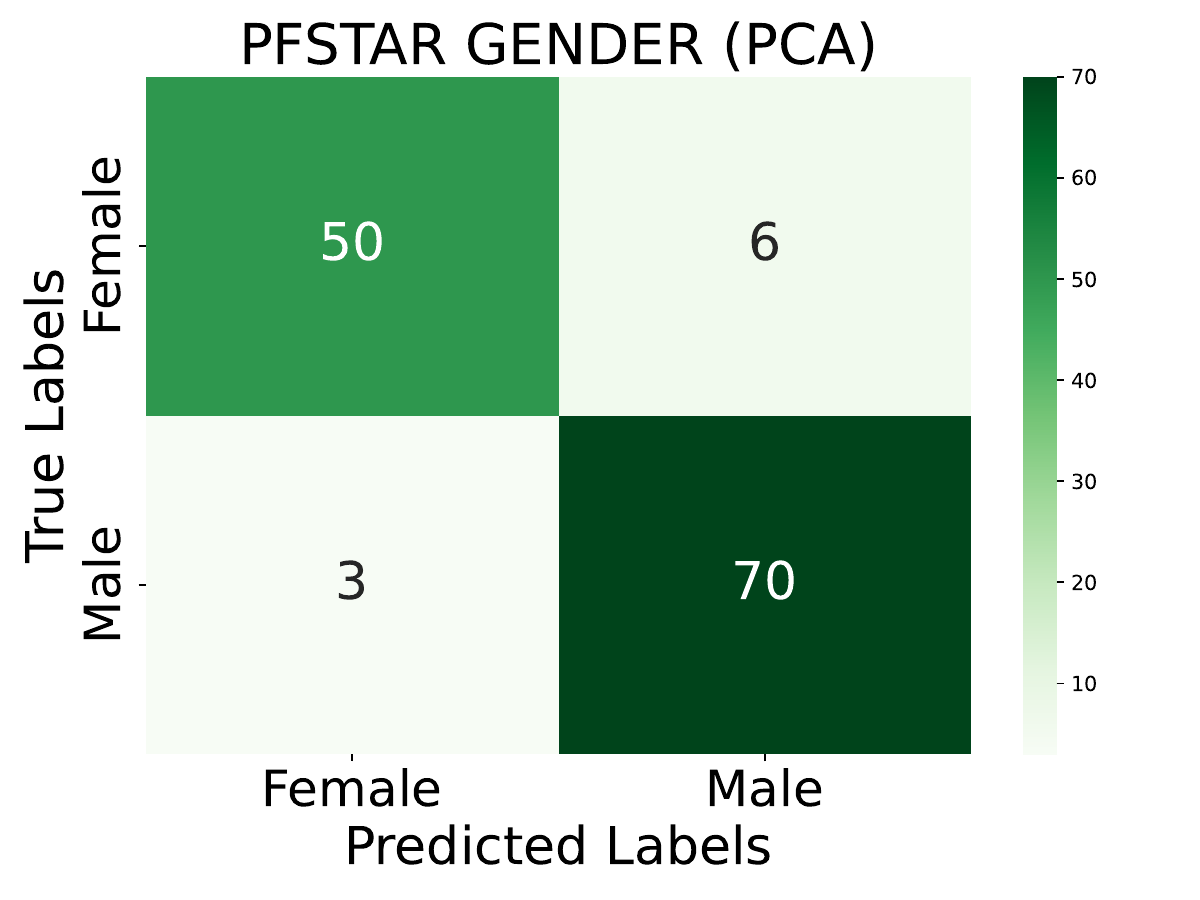}}
    \subfloat[]
    {\includegraphics[width=0.23\textwidth]{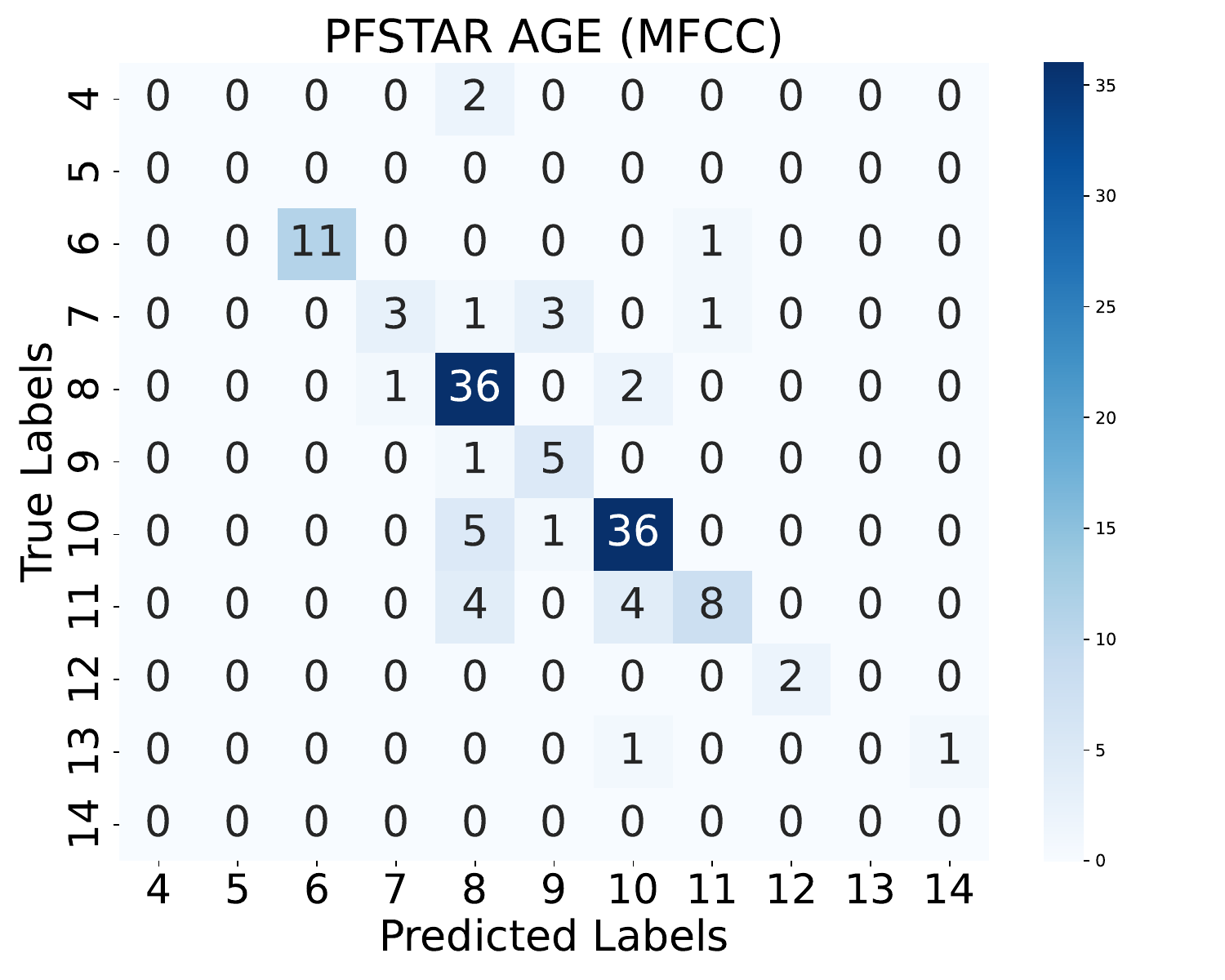}}
    \subfloat[]
    {\includegraphics[width=0.23\textwidth]{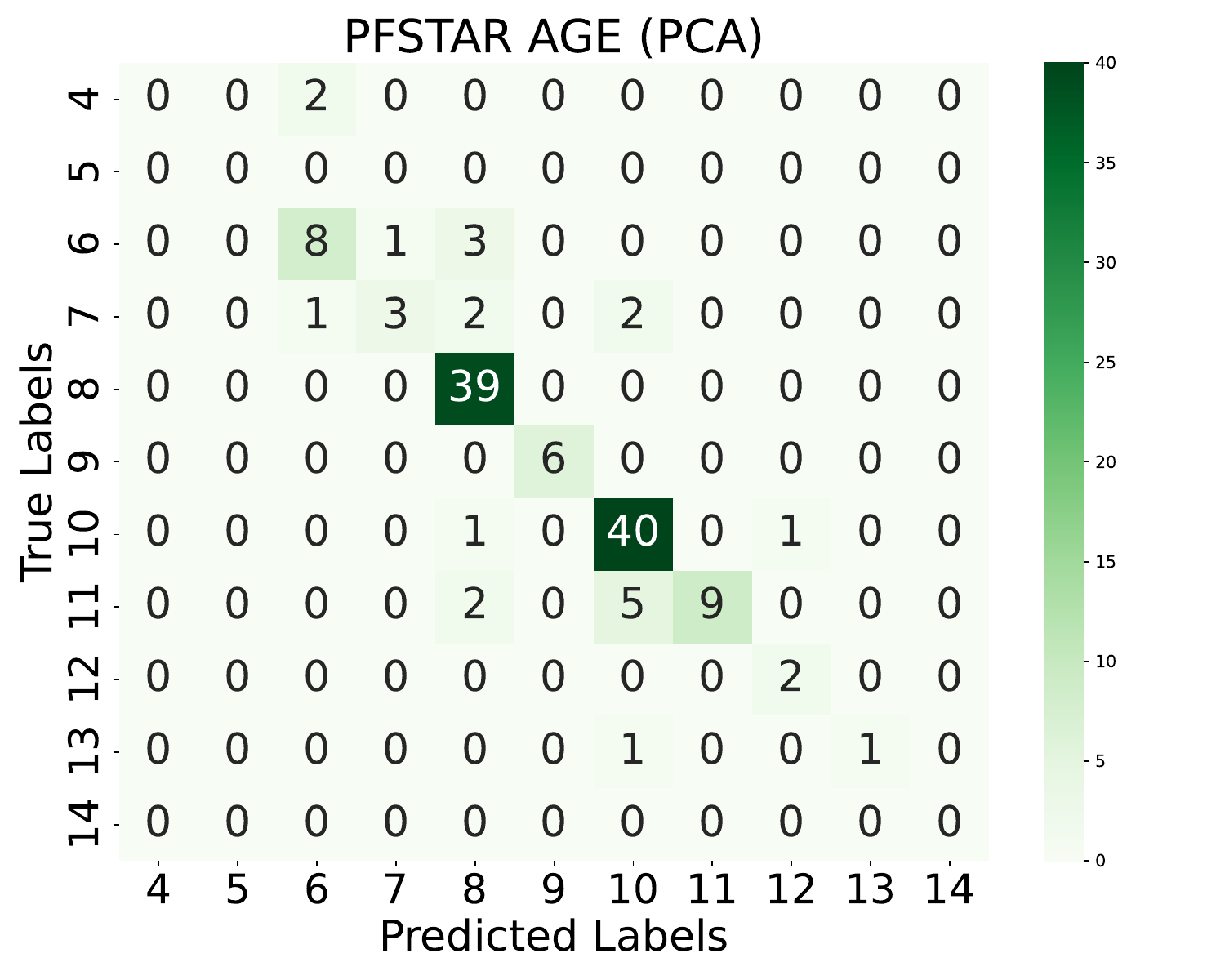}}
    
    \subfloat[]
    {\includegraphics[width=0.23\textwidth]{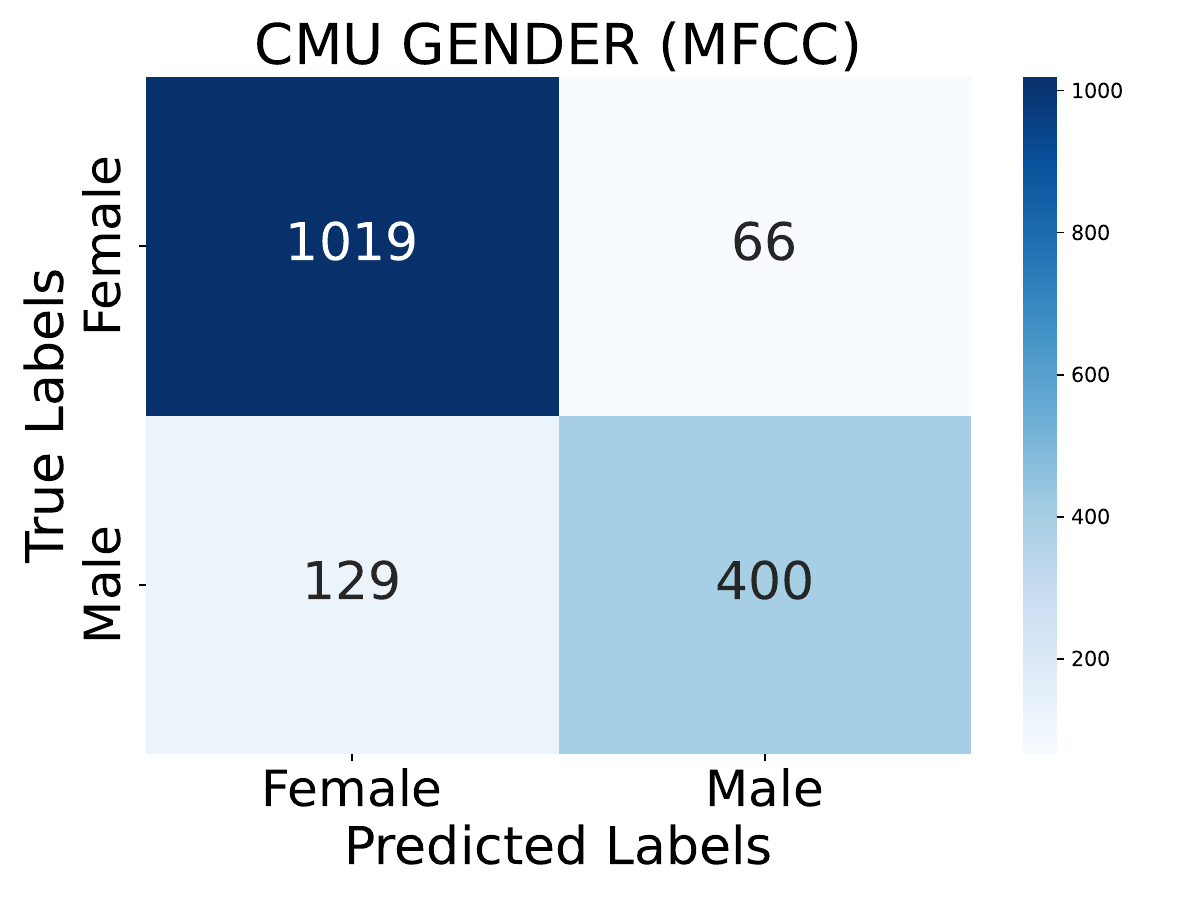}}
    \subfloat[]
    {\includegraphics[width=0.23\textwidth]{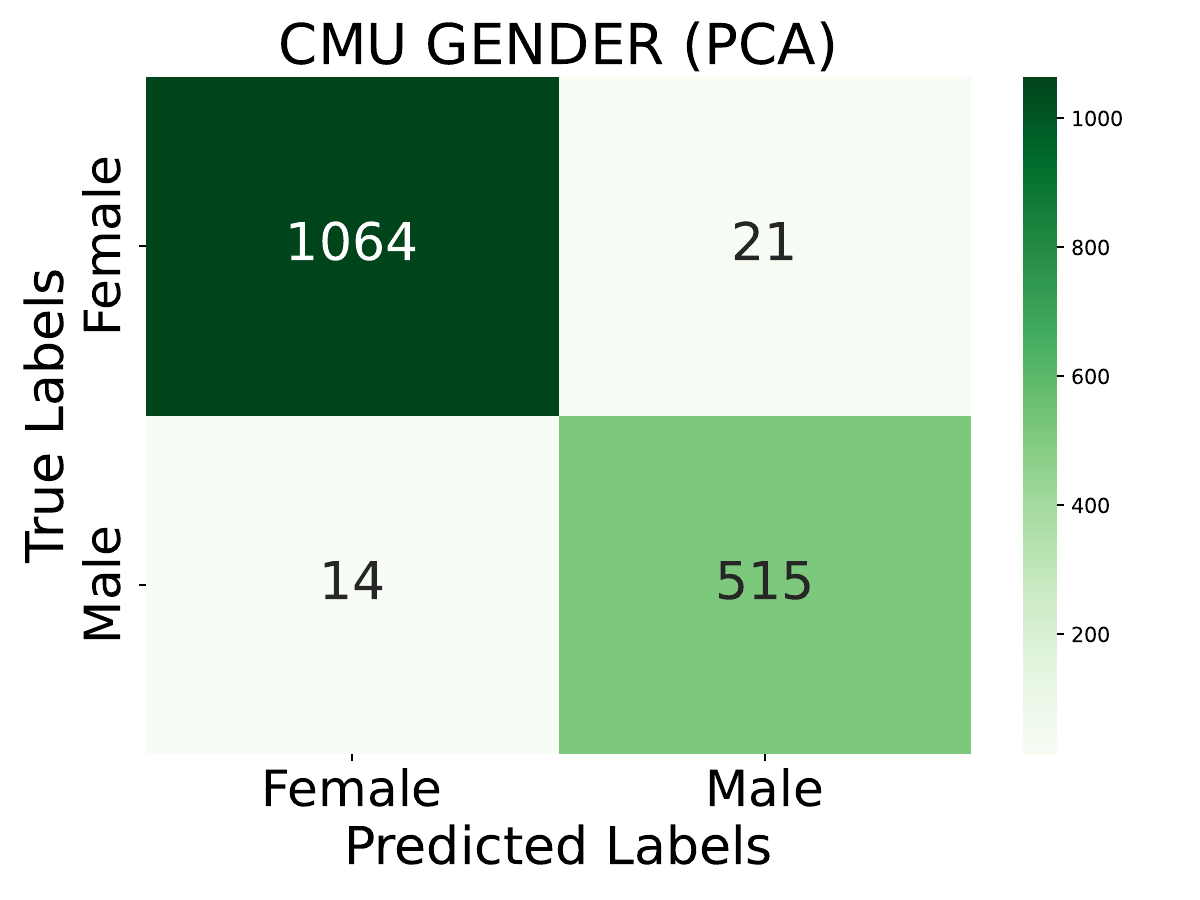}}
    \subfloat[]
    {\includegraphics[width=0.23\textwidth]{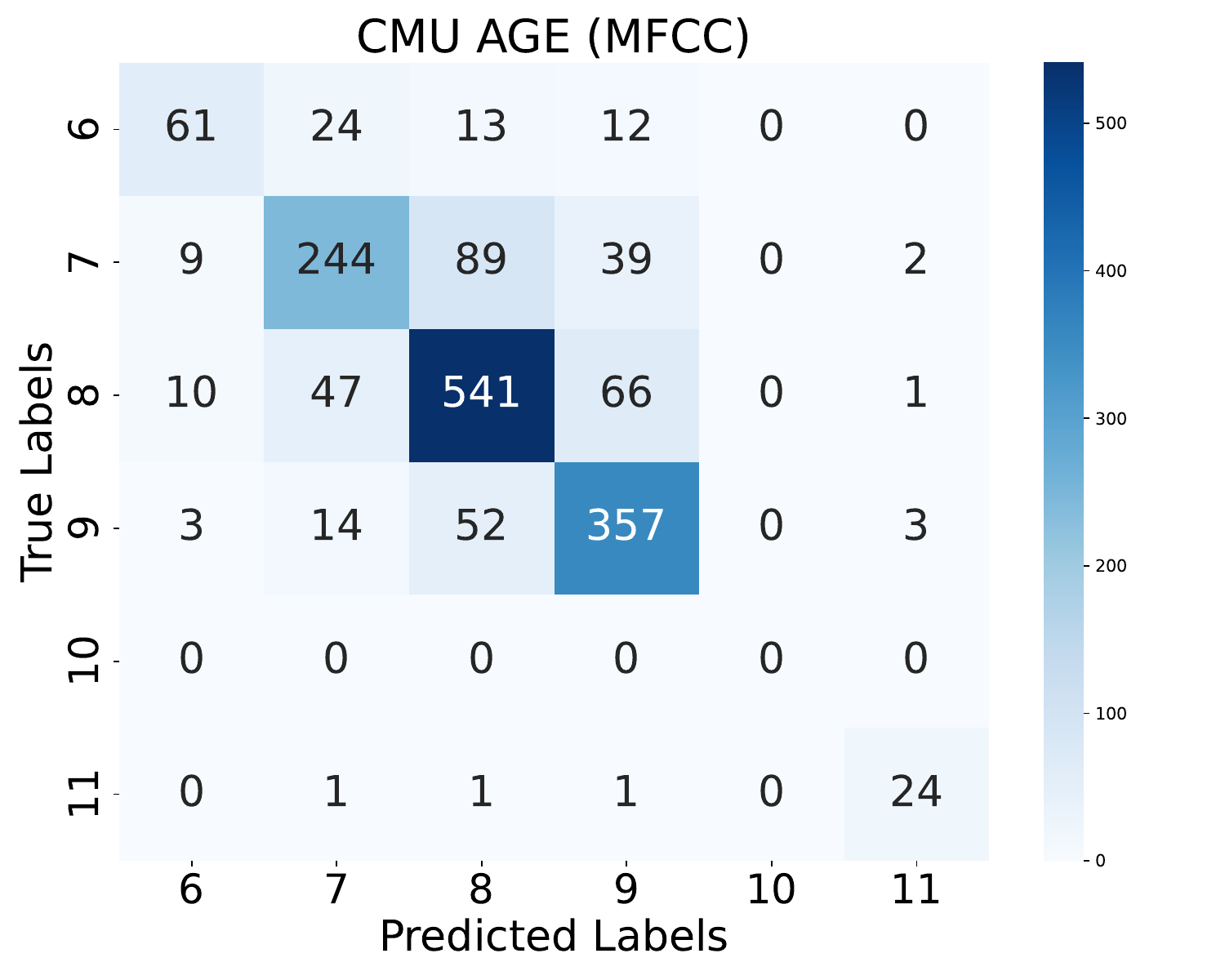}}
    \subfloat[]
    {\includegraphics[width=0.23\textwidth]{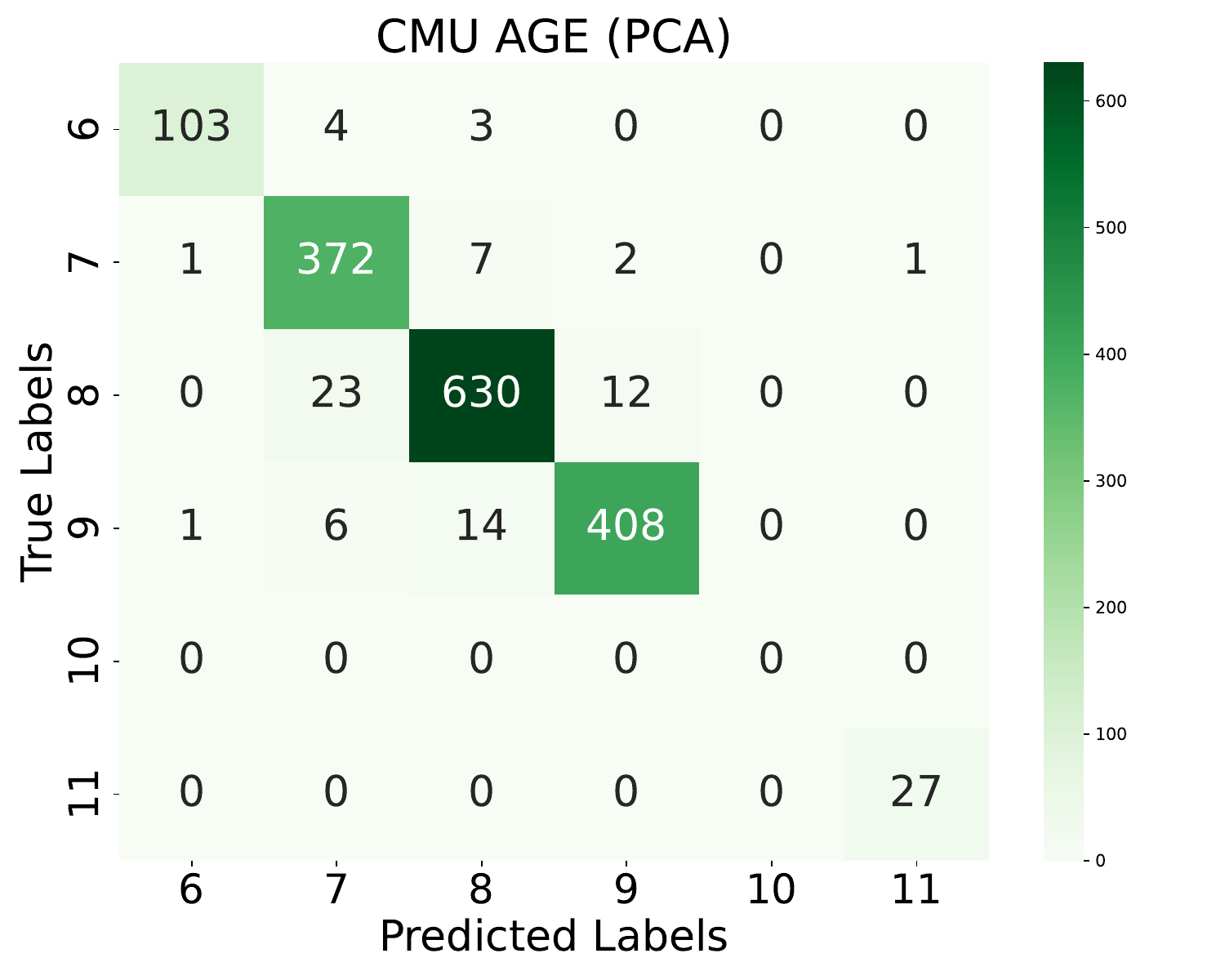}}
    \caption{Comparison of confusion matrices for age and gender classification on the PFSTAR and CMU Kids datasets. Blue matrices correspond to the baseline system using MFCC features, while green matrices represent results from the best-performing SSL model layers after PCA-based dimensionality reduction.}
    \label{fig:confusion_matrices}
\end{figure*}

Overall, the results across both datasets clearly indicate that dimensionality reduction can yield highly effective classifiers for age and gender classification. The substantial reduction in feature dimensions without significant loss of performance suggests that not all extracted features are necessary for these tasks. This highlights the potential for optimizing model efficiency by selecting only the most informative features, thereby improving both classification accuracy and computational efficiency. Consequently, the optimal layer-specific features provide a more efficient solution for age and gender classification, validating the utility of dimensionality reduction in SSL-based models.

To gain deeper insight into the model's behavior, we also analyzed the confusion matrices corresponding to the MFCC features and the best-performing reduced feature set after PCA, as illustrated in Figure~\ref{fig:confusion_matrices}.These matrices help reveal specific patterns of misclassification, particularly among younger age groups, where the model often struggles to clearly separate adjacent age ranges. The confusion matrix for the original MFCC features revealed higher misclassification rates, especially in younger age groups. In contrast, the confusion matrix for the best PCA-reduced feature set indicated fewer misclassifications, emphasizing the robustness of SSL features and demonstrating that the reduced feature set focuses more on the most informative aspects of the data. These confusion matrices further support that, after dimensionality reduction, the model's ability to correctly classify both age and gender categories was significantly improved, confirming the effectiveness of dimensionality reduction in optimizing classification performance for this task.


\begin{table}[h]
\centering
\caption{Age group classification results on the PFSTAR dataset, with the best-performing SSL model (HuBERT) after PCA-based dimensionality reduction.}
\label{tab:age_group_pfstar}
\begin{tabular}{|c|c|c|c|c|c|}
\hline
\textbf{Age Group} & {\textbf{A (\%)}} & {\textbf{P}} & {\textbf{R}} & {\textbf{F1}} & \textbf{Overall Accuracy (\%)} \\
\hline
4--6   & 85.71 & 1.00 & 0.86 & 0.92  & \multirow{3}{*}{93.80} \\
7--9   & 100.00 & 0.88 & 1.00 & 0.94 &  \\
10--13 & 90.32 & 0.98 & 0.90 & 0.94 &  \\
\hline
\end{tabular}
\end{table}

\begin{table}[h]
\centering
\caption{Age group classification results on the CMU Kids dataset, with the best-performing SSL model (HuBERT) after PCA-based dimensionality reduction.}
\label{tab:age_group_cmu}
\begin{tabular}{|c|c|c|c|c|c|}
\hline
\textbf{Age Group} & {\textbf{A (\%)}} & {\textbf{P}} & {\textbf{R}} & {\textbf{F1}} & \textbf{Overall Accuracy (\%)} \\
\hline
6--8   & 99.05 & 0.99 & 0.99 & 0.99  & \multirow{2}{*}{98.57} \\
9--11  & 97.37 & 0.98 & 0.97 & 0.97  &  \\
\hline
\end{tabular}
\end{table}

\subsection{Age Group-wise Analysis}
\label{subsec:age_group_analysis}
In the initial experiments, the classification task was defined at the level of individual ages, resulting in 11 classes for the PFSTAR dataset (ages 4–14) and 5 classes for the CMU Kids dataset (ages 6–11, excluding age 10). While this approach provides detailed insight into model performance, it can be affected by large intra-class variability, especially among younger children whose speech characteristics change rapidly from year to year.

To investigate performance under a more generalized setting, we collapsed the year-wise labels into broader age groups. For PFSTAR, ages were grouped into three categories: 4–6, 7–9, and 10–14 years. For CMU Kids, ages were divided into two categories: 6–8 and 9–11 years. This grouping reflects distinct stages of speech development and helps balance the number of samples per class.

Tables~\ref{tab:age_group_pfstar} and~\ref{tab:age_group_cmu} show the classification results for these age groups, with the best-performing SSL model features after PCA dimensionality reduction. For PFSTAR, overall age classification accuracy increased from 89.15\% to 93.80\%. However, the youngest group remained the most challenging with an accuracy of 85.71\%. The middle and older groups both achieved an F1-score of 0.94 and accuracies over 90\%. For CMU Kids, performance was strong across both age groups, with overall accuracy rising from 97.34\% to 98.57\%. The younger group (6–8 years) reached an accuracy of 99.05\% and F1-score of 0.99, and the older group (9–11 years) reached accuracy of 97.37\% and F1-score of 0.97, reflecting how the smaller age range and controlled recording conditions facilitated more consistent predictions.

These results demonstrate that grouping ages into broader ranges simplifies the classification task and aligns well with natural stages in children’s speech development. Younger children remain more difficult to classify due to greater variability in their speech, but age grouping reduces this variation, allowing the model to more clearly separate classes and improve accuracy compared to predicting exact ages.



\begin{table}[h]
\centering
\caption{Impact of speech segment duration on classification performance for the PFSTAR dataset: comparison of 1-second, 2-second, and 3-second (i.e., 1s, 2s, 3s) chunks using MFCC baseline features and top-performing SSL model (HuBERT) features for both age and gender classification.}
\label{tab:chunk_pfstar}
    \resizebox{10cm}{!}{
\begin{tabular}{|c|c|c|c|c|c|c|}
\hline
\multicolumn{7}{|c|}{\textbf{PFSTAR Age Classification}} \\
\hline
\textbf{Model} & \textbf{Layer} & \textbf{Duration} & \textbf{A (\%)} & \textbf{P} & \textbf{R} & \textbf{F1} \\
\hline
\multirow{3}{*}{MFCC} 
& \multirow{3}{*}{-} & 1s & 21.71 &  0.20&  0.16&  0.12\\
&  & 2s & 27.91 &  0.26&  0.17&  0.14\\
&  & 3s & 18.6 &  0.20&  0.15&  0.13\\
\hline
\multirow{3}{*}{HuBERT} 
& \multirow{3}{*}{3} & 1s & 64.57 & 0.71 & 0.65 & 0.64 \\
&  & 2s & 73.23 & 0.77 & 0.73 & 0.73 \\
&  & 3s & 76.38 & 0.78 & 0.76 & 0.76 \\
\hline

\multicolumn{7}{|c|}{\textbf{PFSTAR Gender Classification}} \\
\hline
\textbf{Model} & \textbf{Layer} & \textbf{Duration} & \textbf{A (\%)} & \textbf{P} & \textbf{R} & \textbf{F1} \\
\hline
\multirow{3}{*}{MFCC} 
& \multirow{3}{*}{-} & 1s & 59.69 &  0.71&  0.54&  0.45\\
&  & 2s & 56.59 &  0.28&  0.50&  0.36\\
&  & 3s & 55.04 &  0.47&  0.49&  0.40\\
\hline
\multirow{3}{*}{Wav2Vec2} 
& \multirow{3}{*}{2} & 1s & 81.40 & 0.82 & 0.81 & 0.81 \\
&  & 2s & 75.19 & 0.78 & 0.75 & 0.75 \\
&  & 3s & 83.72 & 0.84 & 0.84 & 0.84 \\
\hline
\end{tabular}}
\end{table}

\subsection{Short-Segment Classification}
\label{subsec: Audio Segmentation}
Our analysis of segmentation, utilizing the top-performing model and layer configuration previously identified, reveals key trade-offs between speech duration and classification performance. As shown in Tables~\ref{tab:chunk_pfstar} and~\ref{tab:chunk_cmu}, longer segments (3 s) consistently improve accuracy across tasks and datasets. For age classification, HuBERT’s accuracy on PFSTAR increased by 18.3\% (from 64.57\% to 76.38\%) when moving from 1 s to 3 s segments, while Wav2Vec2’s gender classification accuracy improved from 81.40\% to 83.72\%. Similarly, the CMU Kids dataset exhibited the same trend, with both age and gender classification performing better as segment duration increased. These results suggest that longer segments provide more contextual information and help models capture nuanced speech characteristics, such as pitch and tone, which are crucial for accurate classification.

The results also reveal important task-specific differences. Age classification proved more sensitive to segment duration than gender classification. Moreover, CMU Kids’ controlled recordings demonstrated more stable performance across durations, with gender classification achieving 92.44\% accuracy (F1 = 0.92) using only 3-second segments. This indicates that while longer segments generally improve performance, the optimal segment length depends on both the classification task and recording conditions.

Practical applications must balance these findings against real-world constraints. The trade-off between accuracy and real-time processing demands is critical, as shorter segments are often required in embedded or real-time systems, where performance may degrade with reduced segment lengths.

\begin{table}[h]
\centering
\footnotesize
\caption{Impact of speech segment duration on classification performance for the CMU Kids dataset: comparison of 1-second, 2-second, and 3-second (i.e., 1s, 2s, 3s) chunks using MFCC baseline features and top-performing SSL model (HuBERT) features for both age and gender classification.}
\label{tab:chunk_cmu}
    \resizebox{10cm}{!}{
\begin{tabular}{|c|c|c|c|c|c|c|}
\hline
\multicolumn{7}{|c|}{\textbf{CMU Age Classification}} \\
\hline
\textbf{Model} & \textbf{Layer} & \textbf{Duration} & \textbf{A (\%)} & \textbf{P} & \textbf{R} & \textbf{F1} \\
\hline
\multirow{3}{*}{MFCC}
& \multirow{3}{*}{-} & 1s & 51.7 & 0.46  & 0.41  & 0.42  \\
&  & 2s & 61.69 & 0.63  & 0.53  & 0.56 \\
&  & 3s & 71.36 & 0.69 & 0.69 & 0.68 \\
\hline
\multirow{3}{*}{HuBERT}
& \multirow{3}{*}{4} & 1s & 33.77 & 0.36 & 0.28 & 0.28 \\
&  & 2s & 76.99 & 0.79 & 0.77 & 0.77 \\
&  & 3s & 81.21 & 0.82 & 0.81 & 0.80 \\
\hline

\multicolumn{7}{|c|}{\textbf{CMU Gender Classification}} \\
\hline
\textbf{Model} & \textbf{Layer} & \textbf{Duration} & \textbf{A (\%)} & \textbf{P} & \textbf{R} & \textbf{F1} \\
\hline
\multirow{3}{*}{MFCC}
& \multirow{3}{*}{-} & 1s & 72.97 & 0.69  & 0.68  & 0.69  \\
&  & 2s & 79.85 & 0.77  & 0.76 & 0.76 \\
&  & 3s & 83.88 & 0.82 & 0.81 & 0.82 \\
\hline
\multirow{3}{*}{HuBERT}
& \multirow{3}{*}{3} & 1s & 72.41 & 0.71 & 0.72 & 0.70 \\
&  & 2s & 85.99 & 0.88 & 0.86 & 0.85 \\
&  & 3s & 92.44 & 0.93 & 0.92 & 0.92 \\
\hline
\end{tabular}}
\end{table}

{
\subsection{Robustness and Generalization of Layer-wise SSL Representations}

\label{sec:robustness}

The layer-wise analyses presented so far demonstrate that early to mid-level SSL representations encode strong age- and gender-related cues in children’s speech. However, the preceding sections focus on within-corpus layer-wise behavior under fixed evaluation conditions, whereas practical deployment of SSL-based representations requires robustness to domain mismatch, layer selection variability, and amount of data for validation. We therefore conduct a set of complementary analyses to examine the generalization and stability of the observed layer-wise trends. Specifically, we evaluate cross-database transfer, layer aggregation strategies, and k-fold cross-validations.

\subsubsection{Cross-Database Evaluation}

To assess robustness under domain mismatch, we perform cross-database gender classification
experiments between PFSTAR (British English) and CMU Kids (American English). Age
classification is not considered in this setting because the age ranges of PFSTAR (4--14
years) and CMU Kids (6--11 years, excluding age 10) do not align, making age labels
non-comparable across corpora. We report results using MFCC features as a baseline and
HuBERT embeddings extracted from the best-performing layers identified for each dataset
(Layer~6 for PFSTAR and Layer~3 for CMU Kids).

The cross-database results in Tables~\ref{tab:cross-db-gender},
\ref{tab:crossdb-gender-HuBERT3}, and \ref{tab:HuBERT6-gender-crossdb} reveal consistent
trends across both accuracy and F1-score. MFCC-based systems exhibit poor generalization
under cross-corpus conditions. When trained on CMU Kids and evaluated on PFSTAR,
accuracy drops to 51.94\% with an F1-score of 0.52, while training on PFSTAR and testing on
CMU Kids results in a further degradation to 34.01\% accuracy and an F1-score of 0.29.
The pronounced reduction in F1-score indicates a strong imbalance between precision and
recall under corpus mismatch, reflecting differences in accent, recording conditions, and
speaking style.

\begin{table}[!h]
\centering
\caption{Performance metrics — accuracy (A), precision (P), recall (R), and F1 score (F1) for cross-database gender classification using baseline MFCC features, evaluated on the PFSTAR and CMU Kids datasets.}
\label{tab:cross-db-gender}
\begin{tabular}{|l|c|c|c|c|}
\hline
\textbf{Train → Test} & \textbf{A (\%)} & \textbf{Prec.} & \textbf{Recall} & \textbf{F1} \\
\hline
CMU → PFSTAR & 51.94 & 0.5297 & 0.5296 & 0.5194 \\
PFSTAR → CMU & 34.01 & 0.4537 & 0.4860 & 0.2939 \\
\hline
\end{tabular}
\end{table}

In contrast, HuBERT embeddings substantially improve cross-database performance.
Using HuBERT Layer~3, which is optimal for CMU Kids, training on PFSTAR and testing on
CMU Kids achieves 63.26\% accuracy with an F1-score of 0.62, while training on CMU Kids
and evaluating on PFSTAR yields 54.26\% accuracy and an F1-score of 0.51. These results
demonstrate that SSL representations provide more balanced precision--recall trade-offs and
generalize more effectively across datasets than MFCC features.

\begin{table}[!h]
\centering
\caption{Performance metrics — accuracy (A), precision (P), recall (R), and F1 score (F1) for cross-database gender classification using HuBERT Layer~3 embeddings, evaluated on the PFSTAR and CMU Kids datasets.}
\label{tab:crossdb-gender-HuBERT3}
\begin{tabular}{|l|c|c|c|c|}
\hline
\textbf{Train → Test} &
\textbf{A (\%)} &
\textbf{Prec.} &
\textbf{Recall} &
\textbf{F1} \\
\hline
CMU → PFSTAR  
& 54.26 & 0.6387 & 0.5855 & 0.5126 \\
PFSTAR → CMU  
& 63.26 & 0.6337 & 0.6517 & 0.6218 \\
\hline
\end{tabular}
\end{table}

A similar trend is observed for HuBERT Layer~6, identified as the optimal layer for PFSTAR.
As shown in Table~\ref{tab:HuBERT6-gender-crossdb}, training on PFSTAR and testing on
CMU Kids yields 50.25\% accuracy with an F1-score of 0.50, while training on CMU Kids and
testing on PFSTAR results in 51.16\% accuracy and an F1-score of 0.45. Although the
absolute accuracy gains at this layer are more modest than those observed with Layer~3, the
consistently higher F1-scores relative to MFCC baselines indicate more stable and balanced
classification behavior under cross-corpus conditions.

\begin{table}[!h]
\centering
\caption{Performance metrics — accuracy (A), precision (P), recall (R), and F1 score (F1) for cross-database gender classification using HuBERT Layer~6 embeddings, evaluated on the PFSTAR and CMU Kids datasets.}
\label{tab:HuBERT6-gender-crossdb}
\begin{tabular}{|l|c|c|c|c|}
\hline
\textbf{Train} → \textbf{Test} &
\textbf{A (\%)} &
\textbf{Prec.} & \textbf{Recall} & \textbf{F1} \\
\hline
PFSTAR → CMU & 50.25 & 0.6248 & 0.6038 & 0.4988 \\
CMU → PFSTAR & 51.16 & 0.6634 & 0.5643 & 0.4537 \\
\hline
\end{tabular}
\end{table}

\subsubsection{Layer Aggregation}
While the preceding analyses identify task-optimal single layers, selecting a single layer
may be sensitive to dataset or task variations. To reduce this sensitivity and evaluate
whether more stable representations can be obtained, we examine simple, non-trainable
layer aggregation strategies using HuBERT, which consistently yielded the strongest
single-layer performance. Specifically, two aggregation strategies are evaluated. First, a generic early-layer aggregation is performed by computing the mean of HuBERT Layers~0--4, motivated by prior work showing that early SSL layers capture complementary acoustic cues. Second, a task-informed aggregation strategy is employed, where a small set of top-performing layers identified from the layer-wise analysis is selected and combined. For this setting, we compute mean aggregation over the best five to best two layers, and additionally evaluate concatenation of the best two layers. All aggregation operations are applied post hoc to extracted features and do not involve any additional training or fine-tuning.

Tables~\ref{tab:pfstar-agg} and~\ref{tab:cmu-agg} summarize the aggregation results for PFSTAR and CMU Kids, respectively. For age classification, aggregation consistently improves performance over the best single layer across both datasets. On PFSTAR, accuracy increases from 87.40\% for the best single layer to 92.12\% using mean aggregation
over the best three layers, with a corresponding improvement in F1-score from 0.87 to 0.91. Similar trends are observed on CMU Kids, where mean aggregation over the best five layers achieves the highest accuracy of 99.25\% and an F1-score of 0.99, outperforming the best single-layer baseline.

\begin{table}[!t]
\centering
\caption{Performance metrics — accuracy (A), precision (P), recall (R), and F1 score (F1) for layer-wise feature aggregation strategies in age and gender classification on the PFSTAR dataset.}
\label{tab:pfstar-agg}
\begin{tabular}{lccccc}
\hline
\multicolumn{6}{c}{\textbf{Age Classification}} \\
\hline
\textbf{Method} & \textbf{Dim} & \textbf{A (\%)} & \textbf{Prec} & \textbf{Rec} & \textbf{F1} \\
\hline
Best Single Layer      & 1024 & 87.40 & 0.88 & 0.87 & 0.87 \\
\hline
Mean (Layers 0--4)     & 1024 & 89.76 & 0.90 & 0.89 & 0.89 \\
Mean (Best 5 Layers)   & 1024 & 87.40 & 0.89 & 0.87 & 0.87 \\
Mean (Best 4 Layers)   & 1024 & 88.97 & 0.90 & 0.89 & 0.89 \\
Mean (Best 3 Layers)   & 1024 & 92.12 & 0.92 & 0.92 & 0.91 \\
Mean (Best 2 Layers)   & 1024 & 91.33 & 0.92 & 0.91 & 0.91 \\
Concat (Best 2 Layers) & 2048 & 92.91 & 0.93 & 0.93 & 0.92 \\
\hline
\multicolumn{6}{c}{\textbf{Gender Classification}} \\
\hline
\textbf{Method} & \textbf{Dim} & \textbf{Acc (\%)} & \textbf{Prec} & \textbf{Rec} & \textbf{F1} \\
\hline
Best Single Layer      & 1024 & 94.57 & 0.96 & 0.94 & 0.93 \\
Mean (Layers 0--4)     & 1024 & 91.47 & 0.91 & 0.91 & 0.91 \\
Mean (Best 5 Layers)   & 1024 & 89.14 & 0.89 & 0.89 & 0.89 \\
Mean (Best 4 Layers)   & 1024 & 94.57 & 0.94 & 0.94 & 0.94 \\
Mean (Best 3 Layers)   & 1024 & 89.92 & 0.90 & 0.90 & 0.90 \\
Mean (Best 2 Layers)   & 1024 & 92.24 & 0.92 & 0.92 & 0.92 \\
Concat (Best 2 Layers) & 2048 & 92.24 & 0.92 & 0.92 & 0.92 \\
\hline
\end{tabular}
\vspace{-30pt}
\end{table}

\begin{table}[!b]
\centering
\caption{Performance metrics — accuracy (A), precision (P), recall (R), and F1 score (F1) for layer-wise feature aggregation strategies in age and gender classification on the CMU Kids dataset.}
\label{tab:cmu-agg}
\begin{tabular}{lccccc}
\hline
\multicolumn{6}{c}{\textbf{Age Classification}} \\
\hline
\textbf{Method} & \textbf{Dim} & \textbf{Acc (\%)} & \textbf{Prec} & \textbf{Rec} & \textbf{F1} \\
\hline
Best Single Layer      & 1024 & 97.33 & 0.97 & 0.97 & 0.97 \\
\hline
Mean (Layers 0--4)     & 1024 & 99.07 & 0.99 & 0.99 & 0.99 \\
Mean (Best 5 Layers)   & 1024 & 99.25 & 0.99 & 0.99 & 0.99 \\
Mean (Best 4 Layers)   & 1024 & 99.19 & 0.99 & 0.99 & 0.99 \\
Mean (Best 3 Layers)   & 1024 & 99.00 & 0.99 & 0.99 & 0.99 \\
Mean (Best 2 Layers)   & 1024 & 99.13 & 0.99 & 0.99 & 0.99 \\
Concat (Best 2 Layers) & 2048 & 98.94 & 0.99 & 0.99 & 0.99 \\
\hline
\multicolumn{6}{c}{\textbf{Gender Classification}} \\
\hline
\textbf{Method} & \textbf{Dim} & \textbf{Acc (\%)} & \textbf{Prec} & \textbf{Rec} & \textbf{F1} \\
\hline
Best Single Layer      & 1024 & 98.00 & 0.98 & 0.98 & 0.98 \\
Mean (Layers 0--4)     & 1024 & 99.13 & 0.99 & 0.99 & 0.99 \\
Mean (Best 5 Layers)   & 1024 & 99.50 & 0.99 & 0.99 & 0.99 \\
Mean (Best 4 Layers)   & 1024 & 99.31 & 0.99 & 0.99 & 0.99 \\
Mean (Best 3 Layers)   & 1024 & 99.31 & 0.99 & 0.99 & 0.99 \\
Mean (Best 2 Layers)   & 1024 & 99.38 & 0.99 & 0.99 & 0.99 \\
Concat (Best 2 Layers) & 2048 & 98.88 & 0.99 & 0.99 & 0.99 \\
\hline
\end{tabular}
\end{table}

For gender classification, single-layer performance is already strong, particularly on
CMU Kids. Nevertheless, aggregation yields stable and competitive results. On PFSTAR,
mean aggregation over the best four layers matches the best single-layer accuracy
(94.57\%) while maintaining the same F1-score. On CMU Kids, aggregation further improves
performance from 98.00\% to 99.50\% using mean aggregation over the best five layers,
with near-perfect precision and recall across aggregation strategies.

Across both datasets and tasks, mean aggregation over a small number of informative layers
(typically best three to best five) consistently emerges as an effective and robust
alternative to single-layer selection. In contrast, concatenation occasionally yields
marginal gains but at the cost of increased feature dimensionality, without providing a
consistent advantage. These results indicate that age- and gender-related cues are
distributed across multiple early-to-middle SSL layers and can be captured more reliably
through simple aggregation strategies.

\subsubsection{Cross-Validation}
To assess the reliability of the reported results, we perform
speaker-wise $k$-fold cross-validation on both datasets. For PFSTAR, a 7-fold protocol
is adopted to ensure adequate representation of age and gender classes across splits,
while for CMU Kids, 3-fold cross-validation is sufficient due to its more balanced
structure. In all cases, performance is reported as mean~$\pm$~standard deviation
across folds.

The MFCC baseline results under cross-validation, summarized in
Tables~\ref{tab:pfstar-baseline} and~\ref{tab:cmu-baseline}, highlight clear differences
between the two corpora. PFSTAR exhibits substantially higher variability across folds,
which is consistent with its wide age range and uneven age distribution. In contrast,
CMU Kids shows higher and more stable baseline performance, reflecting its narrower and
more balanced age structure.

\begin{table}[!h]
\centering
\caption{Performance metrics — accuracy (A), precision (P), recall (R), and F1 score (F1) for baseline MFCC feature based age and gender classification on the PFSTAR dataset using 7-fold cross-validation.}
\label{tab:pfstar-baseline}
\begin{tabular}{|c|c|c|c|c|}
\hline
\textbf{Task} &
\textbf{A (\%)} &
\textbf{P} &
\textbf{R} &
\textbf{F1} \\
\hline
Age &
$48.42 \pm 4.50$ &
0.47 &
0.48 &
0.47 \\
\hline
Gender &
$66.71 \pm 8.55$ &
0.66 &
0.66 &
0.66 \\
\hline
\end{tabular}
\end{table}

\begin{table}[!h]
\centering
\caption{Performance metrics — accuracy (A), precision (P), recall (R), and F1 score (F1) for baseline MFCC feature based age and gender classification on the CMU Kids dataset using 3-fold cross-validation.}
\label{tab:cmu-baseline}
\begin{tabular}{|c|c|c|c|c|}
\hline
\textbf{Task} &
\textbf{A (\%)} &
\textbf{P} &
\textbf{R} &
\textbf{F1} \\
\hline
Age &
$87.50 \pm 0.66$ &
0.89 &
0.89 &
0.89 \\
\hline
Gender &
$92.75 \pm 0.16$ &
0.92 &
0.92 &
0.92 \\
\hline
\end{tabular}
\end{table}

To evaluate the robustness of SSL representations under this setting, we repeat the
cross-validation experiments using HuBERT embeddings extracted from the best-performing
layer for each task and dataset. As shown in
Tables~\ref{tab:pfstar-age-gender} and~\ref{tab:cmukids-age-gender}, SSL features
consistently outperform MFCCs across all folds. On PFSTAR, HuBERT improves age
classification accuracy from $48.42\%\pm4.50$ to $55.75\%\pm8.04$ and gender classification
from $66.71\%\pm8.55$ to $87.32\%\pm7.35$. On CMU Kids, where the dataset is more balanced,
the SSL-based classifier achieves near-ceiling performance with minimal variation across
folds ($96.79\%\pm0.72$ for age and $97.80\%\pm0.18$ for gender).

\begin{table}[!h]
\centering
\caption{Performance metrics — accuracy (A), precision (P), recall (R), and F1 score (F1) for age and gender classification on the PFSTAR dataset using HuBERT Layer~3 (age) and Layer~6 (gender) embeddings.}
\label{tab:pfstar-age-gender}
\begin{tabular}{|c|c|c|c|c|c|}
\hline
\textbf{Task} & \textbf{Layer} &
\textbf{A (\%)} & \textbf{P} & \textbf{R} & \textbf{F1} \\
\hline
Age & 3 & $55.75 \pm 8.04$ & 0.53 & 0.55 & 0.54 \\
\hline
Gender & 6 & $87.32 \pm 7.35$ & 0.87 & 0.87 & 0.87 \\
\hline
\end{tabular}
\end{table}

\begin{table}[!h]
\centering
\caption{Performance metrics — accuracy (A), precision (P), recall (R), and F1 score (F1) for age and gender classification on the CMU Kids dataset using HuBERT Layer~4 (age) and Layer~3 (gender) embeddings.}
\label{tab:cmukids-age-gender}
\begin{tabular}{c|c|c|c|c|c}
\hline
\textbf{Task} & \textbf{Layer} &
\textbf{Acc (\%)} &
\textbf{Prec.} & \textbf{Recall} & \textbf{F1} \\
\hline
Age & 4 & $96.79 \pm 0.72$ & 0.97 & 0.97 & 0.97 \\
\hline
Gender & 3 & $97.80 \pm 0.18$ & 0.97 & 0.97 & 0.97 \\
\hline
\end{tabular}
\end{table}

These results confirm that the observed improvements obtained with SSL representations
are stable across data partitions and are not artifacts of a particular train--test
split. The higher variance observed on PFSTAR reflects intrinsic dataset characteristics
rather than instability in the evaluation protocol.

}

\section{Conclusion}
\label{sec:conclusion}

This work examined age and gender classification in children’s speech using SSL models, with a particular focus on understanding how speaker-related information is encoded across different representational layers. Children’s speech poses unique challenges due to ongoing physiological development, high acoustic variability, and limited availability of annotated data. By analyzing layer-wise features from four widely used SSL architectures — Wav2Vec2, HuBERT, Data2Vec, and WavLM, on the PFSTAR and CMU Kids datasets, this study provides a detailed and interpretable view of how SSL representations behave when applied to children’s speech.

The results show that age and gender cues are not evenly distributed across SSL models. Instead, early to mid-level layers consistently capture the most discriminative information,
while deeper layers tend to prioritize higher-level linguistic content at the expense of paralinguistic detail. Among the models considered, HuBERT demonstrates particularly
strong performance for age classification, and both HuBERT and Wav2Vec2 perform competitively for gender classification. These findings highlight that careful layer selection
is crucial when using frozen SSL representations for paralinguistic tasks in children’s speech, and that deeper layers are not necessarily optimal for such applications.

Dimensionality reduction experiments further indicate that SSL feature spaces contain a considerable degree of redundancy. Applying PCA allows feature dimensionality to be reduced substantially while retaining most of the discriminative power. This observation is important for practical deployment, especially in real-time or resource-constrained environments where computational efficiency is a key consideration. In addition, the age group-wise analysis reveals a clear developmental trend: classification performance improves with increasing age, reflecting greater acoustic stability in older children and higher variability among younger speakers.

{Beyond these core analyses, this study places strong emphasis on robustness and generalization. Cross-validation analyses shows that the observed layer-wise trends
and performance gains are stable across different data partitions and are not dependent on a particular train--test split. Layer aggregation experiments demonstrate that combining
information from multiple informative layers can reduce sensitivity to single-layer selection, resulting in more stable and reliable representations. Cross-database evaluation
between PFSTAR and CMU Kids further highlights that SSL representations generalize more effectively than traditional MFCC features across datasets with different accents and recording conditions. Finally, short-segment analysis confirms that meaningful age and gender information can still be extracted from brief speech segments, supporting applications that operate under low-latency or limited-input conditions.}

Taken together, these findings show that frozen SSL representations offer a robust and flexible foundation for age and gender classification in children’s speech. By combining
layer-wise analysis with dimensionality reduction and robustness-oriented evaluation, this work bridges interpretability and practical applicability. The proposed framework provides
clear guidance for selecting and deploying SSL-based features in child-speech scenarios, and it opens avenues for future research on other paralinguistic attributes or lightweight
adaptation strategies under low-resource conditions.

\bibliographystyle{IEEEbib}
\bibliography{arxiv_refs}

@article{koenig2008speech,
  title={Speech production variability in fricatives of children and adults: Results of functional data analysis},
  author={Koenig, Laura L and Lucero, Jorge C and Perlman, Elizabeth},
  journal={The Journal of the Acoustical Society of America},
  volume={124},
  number={5},
  pages={3158--3170},
  year={2008},
  publisher={AIP Publishing}
}

@article{yeung2018difficulties,
  title={On the difficulties of automatic speech recognition for kindergarten-aged children},
  author={Yeung, Gary and Alwan, Abeer},
  journal={Interspeech},
  year={2018}
}

@article{lee1999acoustics,
  title={Acoustics of children’s speech: Developmental changes of temporal and spectral parameters},
  author={Lee, Sungbok and Potamianos, Alexandros and Narayanan, Shrikanth},
  journal={The Journal of the Acoustical Society of America},
  volume={105},
  number={3},
  pages={1455--1468},
  year={1999},
  publisher={Acoustical Society of America}
}

@article{Vorperian2007VowelAS,
  title={Vowel acoustic space development in children: a synthesis of acoustic and anatomic data.},
  author={Houri K. Vorperian and Raymond D. Kent},
  journal={Journal of speech, language, and hearing research : JSLHR},
  year={2007},
  volume={50 6},
  pages={1510-45}
}

@article{disfluency, 
author = {Tran, T. and Tinkler, M. and Yeung, G. and Alwan, A. and Ostendorf, M.}, 
title = {Analysis of disfluency in children’s speech}, 
journal = {Interspeech}, 
year = {2020}, 
doi = {10.21437/interspeech.2020-3037} 
}

@inproceedings{claus2013survey,
  title={A survey about databases of children's speech.},
  author={Claus, Felix and Rosales, Hamurabi Gamboa and Petrick, Rico and Hain, Horst-Udo and Hoffmann, R{\"u}diger},
  booktitle={Interspeech},
  pages={2410--2414},
  year={2013}
}

@article{russell2006pf,
  title={The pf-star british english childrens speech corpus},
  author={Russell, Martin},
  journal={The Speech Ark Limited},
  year={2006},
  publisher={Citeseer}
}

@article{eskenazi1997cmu,
  title={The CMU kids corpus},
  author={Eskenazi, Maxine and Mostow, Jack and Graff, David},
  journal={Linguistic Data Consortium},
  volume={11},
  year={1997},
  publisher={Linguistic Data Consortium Philadelphia}
}

@article{baevski2020wav2vec,
  title={wav2vec 2.0: A framework for self-supervised learning of speech representations},
  author={Baevski, Alexei and Zhou, Yuhao and Mohamed, Abdelrahman and Auli, Michael},
  journal={Advances in neural information processing systems},
  volume={33},
  pages={12449--12460},
  year={2020}
}

@inproceedings{safavi2016speaker,
  title     = {Speaker recognition for children's speech},
  author    = {Saeid Safavi and Maryam Najafian and Abualsoud Hanani and Martin Russell and Peter Jančovič and Michael Carey},
  year      = {2012},
  booktitle = {Interspeech},
  pages     = {1836--1839},
  doi       = {10.21437/Interspeech.2012-401},
  issn      = {2958-1796},
}

@inproceedings{kumari2024role,
  title={Role of Acoustics and Prosodic Features for Children's Age Classification},
  author={Kumari, Vishakha and Sinha, Abhijit and Kathania, Hemant Kumar},
  booktitle={International Conference on Signal Processing and Communications (SPCOM)},
  pages={1--5},
  year={2024},
  organization={IEEE}
}

@article{radha2024automatic,
  title={Automatic speaker and age identification of children from raw speech using sincNet over ERB scale},
  author={Radha, Kodali and Bansal, Mohan and Pachori, Ram Bilas},
  journal={Speech Communication},
  volume={159},
  pages={103069},
  year={2024},
  publisher={Elsevier}
}

@inproceedings{sarma2020children,
  title={Children’s age and gender recognition from raw speech waveform using DNN},
  author={Sarma, Mousmita and Sarma, Kandarpa Kumar and Goel, Nagendra Kumar},
  booktitle={Advances in Intelligent Computing and Communication: Proceedings of ICAC 2019},
  pages={1--9},
  year={2020},
  organization={Springer}
}

@article{Pepino2021EmotionRF,
  title={Emotion Recognition from Speech Using Wav2vec 2.0 Embeddings},
  author={Leonardo Pepino and Pablo Ernesto Riera and Luciana Ferrer},
  journal={Interspeech},
  year={2021},
  volume={abs/2104.03502},
  url={https://api.semanticscholar.org/CorpusID:233181984}
}

@inproceedings{Gao2023TwostageFO,
  title={Two-stage Finetuning of Wav2vec 2.0 for Speech Emotion Recognition with ASR and Gender Pretraining},
  author={Yuan Gao and Chenhui Chu and Tatsuya Kawahara},
  booktitle={Interspeech},
  year={2023},
  url={https://api.semanticscholar.org/CorpusID:260914247}
}

@article{li2013automatic,
  title={Automatic speaker age and gender recognition using acoustic and prosodic level information fusion},
  author={Li, Ming and Han, Kyu J and Narayanan, Shrikanth},
  journal={Computer Speech \& Language},
  volume={27},
  number={1},
  pages={151--167},
  year={2013},
  publisher={Elsevier}
}

@article{sanchez2022age,
  title={Age group classification and gender recognition from speech with temporal convolutional neural networks},
  author={S{\'a}nchez-Hevia, H{\'e}ctor A and Gil-Pita, Roberto and Utrilla-Manso, Manuel and Rosa-Zurera, Manuel},
  journal={Multimedia Tools and Applications},
  volume={81},
  number={3},
  pages={3535--3552},
  year={2022},
  publisher={Springer}
}

@article{kwasny2021gender,
  title={Gender and age estimation methods based on speech using deep neural networks},
  author={Kwasny, Damian and Hemmerling, Daria},
  journal={Sensors},
  volume={21},
  number={14},
  pages={4785},
  year={2021},
  publisher={MDPI}
}

@inproceedings{sinha2024effect,
  title={Effect of Speech Modification on Wav2Vec2 Models for Children Speech Recognition},
  author={Sinha, Abhijit and Singh, Mittul and Kadiri, Sudarsana Reddy and Kurimo, Mikko and Kathania, Hemant Kumar},
  booktitle={International Conference on Signal Processing and Communications (SPCOM)},
  pages={1--5},
  year={2024},
  organization={IEEE}
}

@article{SAFAVI2018141,
title = {Automatic speaker, age-group and gender identification from children’s speech},
journal = {Computer Speech \& Language},
volume = {50},
pages = {141-156},
year = {2018},
issn = {0885-2308},
author = {Saeid Safavi and Martin Russell and Peter Jančovič}
}

@article{10.1121/1.5037614,
    author = {Barreda, Santiago and Assmann, Peter F.},
    title = {Modeling the perception of children's age from speech acoustics},
    journal = {The Journal of the Acoustical Society of America},
    volume = {143},
    number = {5},
    pages = {EL361-EL366},
    year = {2018},
    issn = {0001-4966}
}

@inproceedings{Abhijit_wocci,
  title     = {Layer-Wise Analysis of Self-Supervised Representations for Age and Gender Classification in Children’s Speech},
  author    = {Abhijit Sinha and Harishankar Kumar and Mohit Joshi and Hemant Kumar Kathania and Shrikanth Narayanan and Sudarsana Reddy Kadiri},
  year      = {2025},
  booktitle = {7th Workshop on Child Computer Interaction (WOCCI), Interspeech}
}

@article{NOVOTNY2023104490,
title = {Automated prediction of children's age from voice acoustics},
journal = {Biomedical Signal Processing and Control},
volume = {81},
pages = {104490},
year = {2023},
issn = {1746-8094},
author = {Michal Novotny and Roman Cmejla and Tereza Tykalova}
}

@article{mckechnie2018automated,
  title={Automated speech analysis tools for children’s speech production: A systematic literature review},
  author={McKechnie, Jacqui and Ahmed, Beena and Gutierrez-Osuna, Ricardo and Monroe, Penelope and McCabe, Patricia and Ballard, Kirrie J},
  journal={International journal of speech-language pathology},
  volume={20},
  number={6},
  pages={583--598},
  year={2018},
  publisher={Taylor \& Francis}
}

@inproceedings{bocklet2010age,
  title={Age and gender recognition based on multiple systems-early vs. late fusion.},
  author={Bocklet, Tobias and Stemmer, Georg and Zeissler, Viktor and N{\"o}th, Elmar},
  booktitle={INTERSPEECH},
  pages={2830--2833},
  year={2010}
}

@inproceedings{muller2007combining,
  title={Combining short-term cepstral and long-term pitch features for automatic recognition of speaker age.},
  author={M{\"u}ller, Christian A and Burkhardt, Felix},
  booktitle={Interspeech},
  pages={2277--2280},
  year={2007}
}

@article{meinedo2011age,
  title={Age and gender detection in the I-DASH project},
  author={Meinedo, Hugo and Trancoso, Isabel},
  journal={ACM Transactions on Speech and Language Processing (TSLP)},
  volume={7},
  number={4},
  pages={1--16},
  year={2011},
  publisher={ACM New York, NY, USA}
}

@inproceedings{safavi2014identification,
  title={Identification of age-group from children's speech by computers and humans.},
  author={Safavi, Saeid and Russell, Martin J and Jancovic, Peter},
  booktitle={Interspeech},
  pages={243--247},
  year={2014}
}

@article{kaya2017emotion,
  title={Emotion, age, and gender classification in children’s speech by humans and machines},
  author={Kaya, Heysem and Salah, Albert Ali and Karpov, Alexey and Frolova, Olga and Grigorev, Aleksey and Lyakso, Elena},
  journal={Computer Speech \& Language},
  volume={46},
  pages={268--283},
  year={2017},
  publisher={Elsevier}
}

@article{perez2018children,
  title={Children age and gender classification based on speech using ConvNets},
  author={P{\'e}rez-Espinosa, Humberto and Avila-George, Himer and Mart{\i}nez-Miranda, Juan and Espinosa-Curiel, Ismael and Rodriguez-Jacobo, Josefina and Cruz-Mendoza, Hector A},
  journal={Res. Comput. Sci.},
  volume={147},
  number={4},
  pages={23--35},
  year={2018}
}

@inproceedings{spiegl2009analyzing,
  title={Analyzing features for automatic age estimation on cross-sectional data.},
  author={Spiegl, Werner and Stemmer, Georg and Lasarcyk, Eva and Kolhatkar, Varada and Cassidy, Andrew and Potard, Blaise and Shum, Stephen and Song, Young Chol and Xu, Puyang and Beyerlein, Peter and others},
  booktitle={Interspeech},
  volume={2009},
  pages={2923},
  year={2009}
}

@article{van2012calibration,
  title={Calibration of probabilistic age recognition},
  author={Van Leeuwen, David and Bahari, Mohamad Hasan},
  journal={Proceedings Interspeech 2012},
  pages={502--505},
  year={2012},
  publisher={ISCA-INT SPEECH COMMUNICATION ASSOC}
}

@inproceedings{kitagishi2020speaker,
  title={Speaker age estimation using age-dependent insensitive loss},
  author={Kitagishi, Yuki and Kamiyama, Hosana and Ando, Atsushi and Tawara, Naohiro and Mori, Takeshi and Kobashikawa, Satoshi},
  booktitle={Asia-Pacific Signal and Information Processing Association Annual Summit and Conference (APSIPA ASC)},
  pages={319--324},
  year={2020},
  organization={IEEE}
}

@inproceedings{wav2vec2_speaker,
author = {Fan, Zhiyun and Li, Meng and Zhou, Shiyu and Xu, Bo},
year = {2021},
month = {08},
pages = {1509-1513},
title = {Exploring wav2vec 2.0 on Speaker Verification and Language Identification},
doi = {10.21437/Interspeech.2021-1280}
}

@inproceedings{ssl_emotion,
author = {Morais, Edmilson and Hoory, Ron and Zhu, Weizhong and Gat, Itai and Damasceno, Matheus and Aronowitz, Hagai},
year = {2022},
month = {05},
pages = {6922-6926},
title = {Speech Emotion Recognition Using Self-Supervised Features},
doi = {10.1109/ICASSP43922.2022.9747870}
}

@article{ANIDJAR2024124671,
title = {Harnessing the power of Wav2Vec2 and CNNs for Robust Speaker Identification on the VoxCeleb and LibriSpeech Datasets},
journal = {Expert Systems with Applications},
volume = {255},
pages = {124671},
year = {2024},
issn = {0957-4174},
doi = {https://doi.org/10.1016/j.eswa.2024.124671},
author = {Or Haim Anidjar and Revital Marbel and Roi Yozevitch},
}

@INPROCEEDINGS{9747379,
  author={Zhu, Qiu-Shi and Zhang, Jie and Zhang, Zi-Qiang and Wu, Ming-Hui and Fang, Xin and Dai, Li-Rong},
  booktitle={IEEE International Conference on Acoustics, Speech and Signal Processing (ICASSP)}, 
  title={A Noise-Robust Self-Supervised Pre-Training Model Based Speech Representation Learning for Automatic Speech Recognition}, 
  year={2022},
  volume={},
  number={},
  pages={3174-3178},
  doi={10.1109/ICASSP43922.2022.9747379}}

@inproceedings{novoselov23_interspeech,
  title     = {On the robustness of wav2vec 2.0 based speaker recognition systems},
  author    = {Sergey Novoselov and Galina Lavrentyeva and Anastasia Avdeeva and Vladimir Volokhov and Nikita Khmelev and Artem Akulov and Polina Leonteva},
  year      = {2023},
  booktitle = {Interspeech},
  pages     = {3177--3181},
  doi       = {10.21437/Interspeech.2023-881},
  issn      = {2958-1796},
}

@inproceedings{kitagishi23_interspeech,
  title     = {What are differences? Comparing DNN and Human by Their Performance and Characteristics in Speaker Age Estimation},
  author    = {Yuki Kitagishi and Naohiro Tawara and Atsunori Ogawa and Ryo Masumura and Taichi Asami},
  year      = {2023},
  booktitle = {Interspeech},
  pages     = {1873--1877},
  doi       = {10.21437/Interspeech.2023-1507},
  issn      = {2958-1796},
}

@INPROCEEDINGS{kang2023svldl,
  author={Kang, Zuheng and Wang, Jianzong and Peng, Junqing and Xiao, Jing},
  booktitle={IEEE Spoken Language Technology Workshop (SLT)}, 
  title={SVLDL: Improved Speaker Age Estimation Using Selective Variance Label Distribution Learning}, 
  year={2022},
  volume={},
  number={},
  pages={1037-1044}
}

@article{wav2vec2,
  title={wav2vec 2.0: A framework for self-supervised learning of speech representations},
  author={Baevski, Alexei and Zhou, Yuhao and Mohamed, Abdelrahman and Auli, Michael},
  journal={Advances in neural information processing systems},
  volume={33},
  pages={12449--12460},
  year={2020}
}

@article{hubert,
  title={Hubert: Self-supervised speech representation learning by masked prediction of hidden units},
  author={Hsu, Wei-Ning and Bolte, Benjamin and Tsai, Yao-Hung Hubert and Lakhotia, Kushal and Salakhutdinov, Ruslan and Mohamed, Abdelrahman},
  journal={IEEE/ACM transactions on audio, speech, and language processing},
  volume={29},
  pages={3451--3460},
  year={2021},
  publisher={IEEE}
}

@inproceedings{data2vec,
  title={Data2vec: A general framework for self-supervised learning in speech, vision and language},
  author={Baevski, Alexei and Hsu, Wei-Ning and Xu, Qiantong and Babu, Arun and Gu, Jiatao and Auli, Michael},
  booktitle={International Conference on Machine Learning},
  pages={1298--1312},
  year={2022},
  organization={PMLR}
}

@article{wavlm,
  title={Wavlm: Large-scale self-supervised pre-training for full stack speech processing},
  author={Chen, Sanyuan and Wang, Chengyi and Chen, Zhengyang and Wu, Yu and Liu, Shujie and Chen, Zhuo and Li, Jinyu and Kanda, Naoyuki and Yoshioka, Takuya and Xiao, Xiong and others},
  journal={IEEE Journal of Selected Topics in Signal Processing},
  volume={16},
  number={6},
  pages={1505--1518},
  year={2022},
  publisher={IEEE}
}

@ARTICLE{1,
  author = {A. Potaminaos and S. Narayanan},
  title = {{Robust Recognition of Children Speech}},
  journal = {IEEE Transactions on Speech and Audio Processing},
  year = {2003},
  volume = {11},
  pages = {603-616},
  number = {6},
  month = {November}
}

@INPROCEEDINGS{3,
  author = {H. Singer and S. Sagayama},
  title = {{Pitch dependent phone modelling for {HMM} based speech recognition}},
  booktitle = {Proc. ICASSP},
  year = {1992},
  pages = {273-276},
  month = {March}
}

@INPROCEEDINGS{4,
  author = {X. Shao and B. Milner},
  title = {{Pitch prediction from MFCC vectors for speech
reconstruction}},
  booktitle = {Proc. ICASSP},
  year = {2004},
  pages = {97-100},
  month = {May}
}

@INPROCEEDINGS{6,
  author = {H. Hirsch and D. Pearce},
  title = {{The AURORA experimental framework for
the performance evaluation of speech recognition systems under noisy
conditions}},
  booktitle = {Proc. ASRU},
  year = {2000},
  pages = {181-188}
}

@inproceedings{grosz2022wav2vec2,
  title={Wav2vec2-based paralinguistic systems to recognise vocalised emotions and stuttering},
  author={Gr{\'o}sz, Tam{\'a}s and Porjazovski, Dejan and Getman, Yaroslav and Kadiri, Sudarsana and Kurimo, Mikko},
  booktitle={Proceedings of the 30th ACM International Conference on Multimedia},
  pages={7026--7029},
  year={2022}
}

@inproceedings{tirronen2023utilizing,
  title={Utilizing wav2vec in database-independent voice disorder detection},
  author={Tirronen, Saska and Javanmardi, Farhad and Kodali, Manila and Kadiri, Sudarsana Reddy and Alku, Paavo},
  booktitle={IEEE International Conference on Acoustics, Speech and Signal Processing (ICASSP)},
  pages={1--5},
  year={2023}
}

@article{javanmardi2024pre,
  title={Pre-trained models for detection and severity level classification of dysarthria from speech},
  author={Javanmardi, Farhad and Kadiri, Sudarsana Reddy and Alku, Paavo},
  journal={Speech Communication},
  volume={158},
  pages={103047},
  year={2024},
  publisher={North-Holland}
}

@article{javanmardi2024exploring,
  title={Exploring the impact of fine-tuning the wav2vec2 model in database-independent detection of dysarthric speech},
  author={Javanmardi, Farhad and Kadiri, Sudarsana Reddy and Alku, Paavo},
  journal={IEEE journal of biomedical and health informatics},
  volume={28},
  number={8},
  pages={4951--4962},
  year={2024},
  publisher={IEEE}
}

@article{hsu2021hubert,
  title={Hubert: Self-supervised speech representation learning by masked prediction of hidden units},
  author={Hsu, Wei-Ning and Bolte, Benjamin and Tsai, Yao-Hung Hubert and Lakhotia, Kushal and Salakhutdinov, Ruslan and Mohamed, Abdelrahman},
  journal={IEEE/ACM transactions on audio, speech, and language processing},
  volume={29},
  pages={3451--3460},
  year={2021},
  publisher={IEEE}
}

@inproceedings{baevski2022data2vec,
  title={Data2vec: A general framework for self-supervised learning in speech, vision and language},
  author={Baevski, Alexei and Hsu, Wei-Ning and Xu, Qiantong and Babu, Arun and Gu, Jiatao and Auli, Michael},
  booktitle={International Conference on Machine Learning},
  pages={1298--1312},
  year={2022},
  organization={PMLR}
}

@article{chen2022wavlm,
  title={Wavlm: Large-scale self-supervised pre-training for full stack speech processing},
  author={Chen, Sanyuan and Wang, Chengyi and Chen, Zhengyang and Wu, Yu and Liu, Shujie and Chen, Zhuo and Li, Jinyu and Kanda, Naoyuki and Yoshioka, Takuya and Xiao, Xiong and others},
  journal={IEEE Journal of Selected Topics in Signal Processing},
  volume={16},
  number={6},
  pages={1505--1518},
  year={2022},
  publisher={IEEE}
}

\end{document}